\documentclass[preprint]{aastex631}
\usepackage{colortbl}
\hypersetup{linkcolor=red,citecolor=blue,filecolor=cyan,urlcolor=blue}

\shorttitle{Connection between subsurface layers and surface magnetic activity}
\shortauthors{Baird, Tripathy \& Jain}

\begin{document}

\title{Connection between Sub-surface Layers and Surface Magnetic Activity \\over Multiple Solar Cycles using GONG Observations}

\correspondingauthor{Sushanta Tripathy}
\email{stripathy@nso.edu}

\author[0009-0003-9686-374X]{Mackenzie A. Baird}
\affiliation{Villanova University, 800 Lancaster Ave, Villanova, PA 19085, USA}
\affiliation{National Solar Observatory, REU Program, 3665 Discovery Dr.,  Boulder, CO 80303, USA}

\author[0000-0002-4995-6180]{Sushanta C. Tripathy}
\affiliation{National Solar Observatory, 3665 Discovery Dr.,  Boulder, CO 80303, USA}

\author[0000-0002-1905-1639]{Kiran Jain}
\affiliation{National Solar Observatory, 3665 Discovery Dr.,  Boulder, CO 80303, USA}
\begin{abstract}
We investigate the spatio-temporal evolution of  high-degree acoustic mode frequencies of the Sun and surface magnetic activity over the course of multiple solar cycles, to improve our understanding of the connection between the solar interior and atmosphere.  We focus on high-degree {\it p}-modes due to their ability to characterize conditions in the shear layer just below the solar surface and analyze 22 years of oscillation frequencies obtained from the Global Oscillation Network Group (GONG). 
Considering 10.7 cm radio flux measurements, the sunspot number, and the local magnetic activity index as solar activity proxies, we find strong correlation between the mode frequencies and each activity index. We further investigate the hemispheric asymmetry associated with oscillation frequencies and magnetic activity  proxies, and find that both were dominant in the southern hemisphere during the descending phase of cycle 23,  while in cycle 24 these quantities fluctuated between northern and southern hemispheres. 
Analyzing the frequencies at different latitudes with the progression of solar cycles,  we observe that the variations at mid-latitudes were dominant in the southern hemisphere during the maximum activity period of cycle 24 but the values overlap 
as the cycle advances towards the minimum phase. 
The mode frequencies at the beginning of cycle 25 are found to be dominant in the southern hemisphere following the pattern of magnetic activity. The analysis provides added evidence that the variability in oscillation frequencies are caused by both strong and weak magnetic fields. 
\end{abstract}


\vskip 0.5in
\section{Introduction} \label{sec:intro}
The acoustic oscillations at the surface of the Sun are known to reveal characteristics of the solar interior. Without the ability to physically see the interior, we rely on seismic waves to comprehend the dynamics below the solar atmosphere mainly through the use of {\it p-}mode frequencies. These frequencies are known to vary with the level of surface magnetic activity and synchronize with the solar magnetic cycle. 
The change of oscillation frequencies  in phase with the progression of solar cycle  were first reported by \citet{Woodard1990} and have been confirmed by a large number of studies covering low-degree \citep{chaplin04,sct-basu12,sct-salabert15,jain18a}, intermediate-degree \citep{Woodard1991,tripathy07,jain13b,korzennik23} and  high-degree global mode frequencies  \citep{CRS2008, sct-rhodes11, Tripathy2015}. A few studies, however,  have alluded that the correlation is weak during the low activity phase, for example, during the solar minima preceding cycle 24 \citep{sct-salabert09,Tripathy10,Jain11}, and cycle 25 \citep{Jain22}. These two minima were unusually long and deep. This anomaly was present in both the length and timing of the minima, as determined by helioseismic frequencies, and apparent in the surface magnetic activity proxies. In addition to temporal variation, the frequencies  of high degree-modes from local helioseismic techniques  have also shown high correlation with the spatial distribution of 
surface magnetic fields \citep{sct-hindman00, sct-rhowe04, Tripathy10b,Tripathy2015}. Since there are numerous evidence that point to a strong linear relationship  between the solar activity and the change in oscillation frequencies, it has been argued that the helioseismic frequency shifts can be considered as another proxy of the Sun's activity cycle \citep{Broomhall2015, sct-salabert15}.  

Furthermore, studies with improved and longer data sets indicate that the relationship between the frequency shifts and solar activity is not always simple \citep[see review by][]{Basu2016}. Using intermediate-degree ($7 \le \ell \le 188$)  global mode frequencies, \citet{Jain2009} have shown that the linear correlation between the change in frequencies and the magnetic activity differs from phase to phase in the activity cycle. Similar results were obtained for low-degree  ($\ell \le 3$) mode frequencies  computed from Sun-as-a-Star observations \citep{Broomhall2015}. 

The early study of frequency variations with solar cycle demonstrated that the frequency changes are predominantly a function of frequency and independent of harmonic degree. This led to the perspective that the solar-cycle dependent frequency changes are confined to a thin layer close to the surface and are caused by the magnetic field at the tachocline evolving with solar cycle \citep{robert86}. Other mechanisms such as structural changes in the sub-surface layers due to temperature \citep{kuhn88, foullon10}, modification of the acoustic cavity  \citep{dzi2005}, influence of magnetic atmosphere \citep{jain96} etc. have also been considered \citep[see review by][]{howe2008}. 

Recent studies further suggest that the magnetic field inside the Sun influences the oscillation frequencies in different ways depending on its location and geometry. Analyzing individual multiplet frequencies (frequencies that are functions of $n$, $\ell$, and $m$)  from Global Oscillation Network Group (GONG) over solar cycles 23 and 24, \citet{Jain22} found that 
the minimum period associated with different acoustic modes occur at different epochs e.g., occurring a year earlier in deeper layers as compared to the surface. This led the authors to  conclude that the magnetic fields located at three different locations, i.e., core, tachocline and near-surface shear layers   may   play important roles in modifying the oscillation frequencies.  Based on theoretical models  where a quasi-degenerate perturbation theory was used to  calculate the effect of the toroidal field on the solar oscillation frequencies,  \citet{Kiefer18} suggested that the magnitude of the observed frequency shifts over the entire solar cycle cannot be explained 
by the presence of strong field produced by the dynamo located in the tachocline region alone and conjectured that part of the frequency shifts may be explained by the magnetic fields near the surface. Analysing the GONG global frequencies, \citet{broomhall2017} has compared the depth of the minimum preceding and following solar cycle 23 and advocated that the near-surface magnetic perturbation is responsible for the changes in the frequencies.  
  
 Considering the dominant physical mechanism to be sub-surface magnetic perturbation, a better knowledge of the near-surface layers is required to advance our understanding. This can be achieved through a detailed analysis of accurate and precise frequencies of high-degree modes since the lower turning points of these modes occur very close to the surface.  However, determination of high-degree modes using the techniques of global helioseismology is a challenging task since individual peaks are blended into ridges. 
 In spite of this inherent difficulty, there have been several attempts to compute high-degree oscillation frequencies for different epochs \citep[][and references therein]{Korzennik2013a, reiter2015}.  
 
  In addition to global helioseismology, the local helioseismic technique of ring diagram \citep{Hill1988} also produces frequencies of high-degree modes and are available since mid-2001 from GONG\footnote{https://gong.nso.edu}.  In this paper we analyze these high-degree acoustic mode frequencies as a tool to understand the characteristics of subsurface layers and their connection with the surface magnetic field. This analysis extends our previous study \citep[][hereafter paper I]{Tripathy2015} where we  studied a part of solar cycle 23 and the following minimum.  With the data available for another solar cycle, here we expand the investigation to include cycle 24 and the  minimum between cycles 24 and 25. This allows us to extend the results reported in Paper I to another solar cycle that is weaker compared to cycle 23.  The rest of the paper is organized as follows: We describe the data in Section~2 and present our results in Section~3. We summarize our findings in Section~4.

\section{Data Analysis} \label{sec:floats}
The high-degree mode frequencies used in this study are computed from merged GONG 1-min full-disk Dopplergrams \citep{Harvey96,Harvey98} and cover the period from 2001 July 26  to 2023 March 22. This time range includes approximately the last seven years of cycle 23, complete cycle 24, and the first three years of cycle 25. The mode frequencies are determined through the standard GONG ring-diagram pipeline \citep{Corbard2003}. A good description of the ring-diagram technique can be found in \citet{Hill1988} but for sake of completeness we present a brief outline of the procedure used in this work.  The solar surface in each  Dopplergram is divided into 189 overlapping regions  (tiles hereafter) covering $\pm 52.5^\circ$ in latitude and central meridian distance (CMD), spaced by $7.5^\circ$ in either direction. Each tile covers an area of 15$^{\circ}$ $\times$ 15$^{\circ}$ in heliographic longitude and latitude after apodization, and is tracked for 1664 minutes (hereafter `ring-day') using Snodgrass rotation rate \citep{Snodgrass1984}. The collection of 189 data cubes centered on different latitudes and longitudes is commonly referred as a dense-pack mosaic \citep{sct-haber00}.   Each data cube is passed through a three-dimensional Fourier transform to create a three-dimensional power spectrum with two dimensions in space and one in time. The resulting power spectrum is fitted with a Lorentzian profile described by
\begin{equation}
    P\left(k_x,k_y,\nu\right) = \frac{A\Gamma}{\left(\nu - \nu_0 + k_xU_x + k_yU_y\right)^2 + \Gamma^2} + \frac{b}{k^3} \;    ,
     \label{RD}
\end{equation}
where $P$ is the oscillation power for a wave with a temporal frequency $\nu$. 
The horizontal wave number, $k$, is defined as a sum of  $k_x\, \hat{x} + k_y\, \hat{y}$ where $\hat{x}$ and $\hat{y}$ point in the prograde and northward directions, respectively. 
The other six parameters that are fitted consist of the zonal, $U_x$, and meridional, $U_y$, velocities, the mode's unperturbed central frequency in the rest
frame of the Sun, $\nu_0$, the mode amplitude, $A$, the mode width, $\Gamma$, and the background power, $b$. The fitting algorithm uses the maximum-likelihood procedure described in \citet{anderson90} and provides about 200 modes  for radial order, $n$, between 0 -- 6, and degree, $\ell$, between 180 -- 1200 in the frequency range of 1500 -- 5200~$\mu$Hz (see Figure~\ref{lnu} in Appendix~A to visualize the modes fitted by this technique).  Throughout this study, we will refer to the centers of the tiles when describing  latitude or longitude values.  

\begin{figure}
    \centering
\includegraphics[width=0.75\textwidth,angle=90]{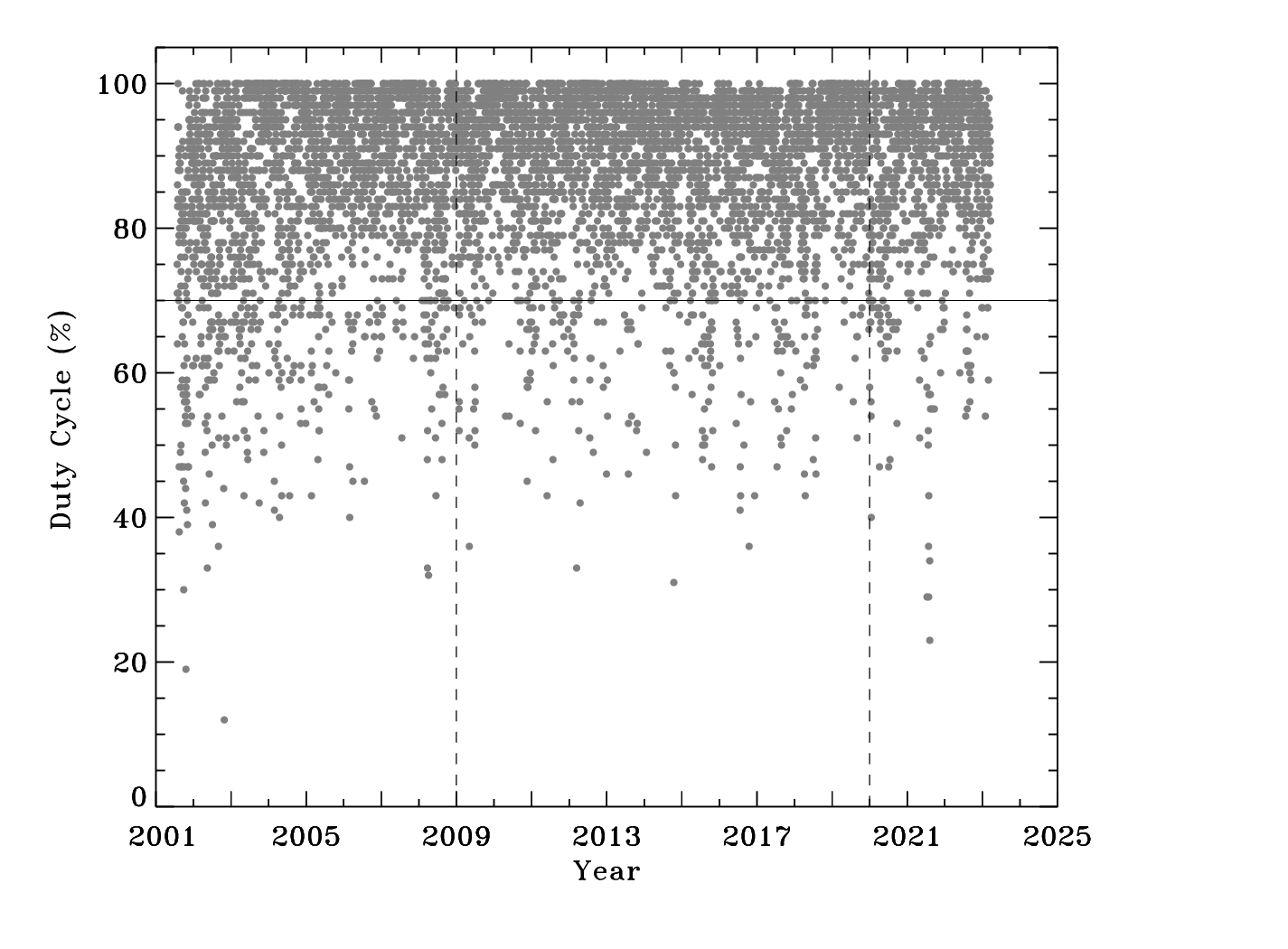} 
\caption{Distribution of duty cycle as a function of time. The data below 70\% duty cycle (horizontal line) are not used in this study. The vertical dashed lines demarcate the solar cycles 23, 24 and 25 based on 13-month smoothed monthly sunspot number (see Section~\ref{sec:activity} for details). \label{dutytime}}
\end{figure}

\begin{figure}
    \centering
\includegraphics[width=0.9\textwidth]{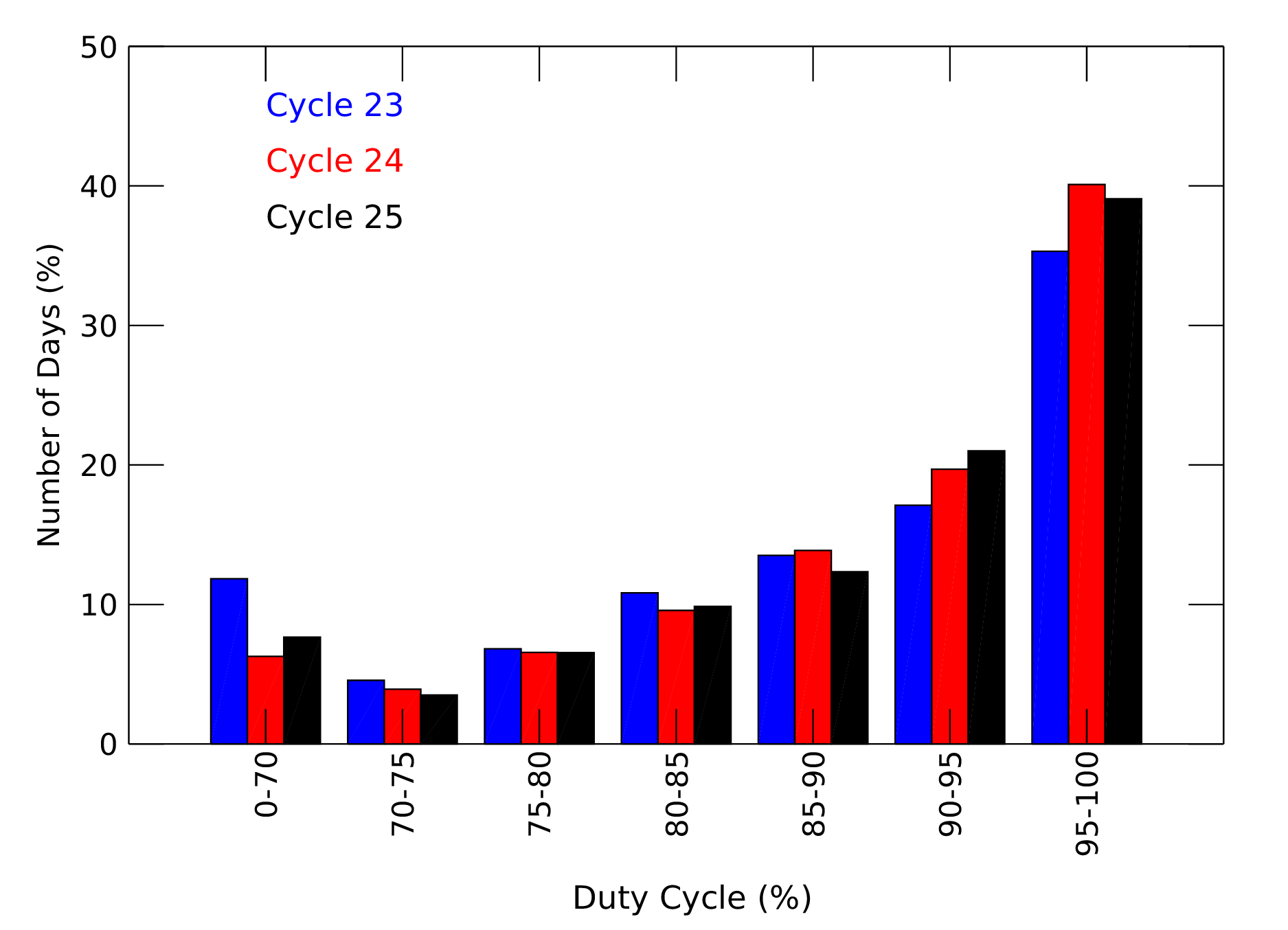} 
\caption{Number of observed ring-days, expressed in percentage, as a function of duty cycle for different solar cycles.  \label{duty}}
\end{figure}

GONG is a ground-based instrument and experiences gaps in observations due to weather conditions and maintenance of instruments \citep{Jain2021}. Figure~\ref{dutytime} shows the distribution of the duty cycle for the analysis period, where the duty cycle is defined as the percentage of good observations out of 1664 minutes. It is clear that the duty cycle varies widely with a mean and median value of 88\% and 92\%, respectively.  Since the mode parameters corresponding to low duty cycle have large uncertainty in fitted values, we exclude the days where the duty cycle is less than $70\%$.   This yielded 6375 usable ring days (92\%) out of 6960 ring-days. We further split the entire data into individual cycles consisting of 2107, 3314, and 954  ring-days for cycles 23, 24, and 25, respectively.  Figure~\ref{duty} shows the number of days as a function of the duty cycle for individual solar cycles in the form of histograms.  We note that cycle 24 has the maximum days with good observations compared to other cycles. 
\subsection{Solar Activity}\label{sec:activity}
Since the mode frequencies are known to vary with solar activity cycle, we compute a quantity known as  ``magnetic activity index'' (MAI) as a measure of solar activity. For each tile, a data cube is generated by tracking and re-mapping line-of-sight magnetograms  in the same fashion as Dopplergrams. The absolute value of each pixel higher than a specified threshold in the data cube are then averaged to compute the MAI. It may be noted that the merged GONG  magnetograms are not available prior to 2007 April,  hence we use  96-minute magnetograms from the Michelson Doppler Imager (MDI) up to 2007 April 9  and GONG  magnetograms sampled every 32 minutes thereafter.  The threshold value for MDI magnetograms is set to 50~G \citep{sct-basu04} and 8.8~G for GONG data based on the estimated noise of the instruments. Since the measured magnetic strength between different instruments differ, we scaled 
the MDI values by 0.45 to generate an uniform set of MAI values. The conversion factor is based on the calculation of \citet{Riley2014} where pixel values were  compared between the line-of-sight magnetograms. 
We further calculate the mean magnetic activity index (MMAI), which represents the spatial average of the MAI over the dense-pack mosaic. Since MAIs do not have measured uncertainties, we use the standard error (the standard deviation of MAI of the dense-pack mosaic divided by the square-root of number of contributing tiles) as the uncertainty in the measurement of MMAI.

We further use three additional surface activity proxies to characterize the solar activity; total and hemispheric sunspot number (TSN, HSN, version 2 provided by the WDC-SILSO\footnote{\url{ https://www.sidc.be/silso/datafiles}},  \citeauthor{clette2014}, \citeyear{clette2014}) and the 10.7 cm radio flux, ($F_{10.7},$)\footnote{\url{https://www.spaceweather.gc.ca/forecast-prevision/solar-solaire/solarflux/sx-5-en.php}}. The latter is a measurement of the integrated emissions at a wavelength of 10.7~cm over the entire solar-disk \citep{Covington69,ermolli14} and is expressed in solar flux units (1\,sfu~=~{\rm{10$^{-22}\,$W\,m$^{-2}$\,Hz$^{-1}$}}). The values of all activity indices are interpolated to the same temporal grid to have the same length as the frequency data points. The activity minimum period is chosen on the basis of the 13-month smoothed monthly sunspot number and occurred in December 2008 and December 2019 for cycle 24 and 25 respectively\footnote{\url{https://www.sidc.be/silso/cyclesminmax}}. Accordingly, we consider 2008 December 31 and 2019 December 31 as the days to distinguish between solar cycles 23 and 24 and 24 and 25, respectively. The two vertical dashed lines in Figures~1, 5, 7-8, 10-15, and 18 indicate the separation between the cycles.    

\section{Results and Discussion}
\subsection{Spatial Variability in Frequencies}

One of the advantages of using ring-diagram technique is that it allows us to  investigate the variation of frequencies at different locations over the solar disk and to find their relation with the local magnetic field on a shorter time scale e.g. one ring-day.  Comparing dense-pack mosaics of frequency shifts and co-temporal magnetograms for a few days, \citet{sct-hindman00}  demonstrated that the active regions appear as locations of large frequency shifts. Subsequent analysis using  data over several years suggested that this conclusion is valid only during periods of high activity but breaks down during the solar minimum \citep{Tripathy10, Tripathy2015}. These findings are similar to those obtained by \citet{Jain2009} for global modes. Here we extend the analysis to compare the spatial variability between different solar cycles by computing the daily changes in frequencies, $\delta\nu_s$, which is defined as the weighted sum of the frequency differences of each mode in a given tile where the weights correspond to square of the fitting uncertainties $\sigma_{n,l}$ following the relation
\begin{equation}
    \delta\nu_{s} = \Sigma_{n,l}\frac{\delta\nu_{n,l}}{\sigma_{n,l}^2} \Bigm/ \Sigma_{n,l}\frac{1}{\sigma_{n,l}^2} \;,
     \label{eq_spatial}
\end{equation}
where the frequency differences of each mode ($\delta\nu_{n,l}$) is calculated with respect to the error-weighted spatial average of the same mode present in the dense-pack mosaic of  each ring-day. Thus, $\delta\nu_s$ represents the frequency shift of individual tiles.

\begin{figure}
\centering
   \includegraphics[width=0.8\textwidth]{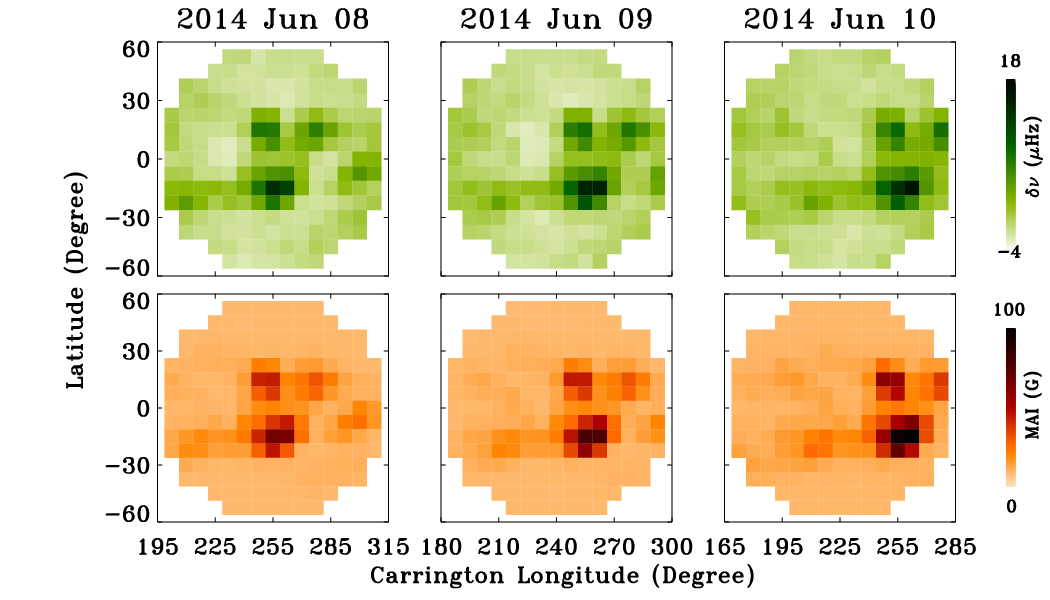}   
   \caption{Spatial distribution of frequency shifts over 189 tiles (top) and coeval MAI (bottom) during high activity period in 2014 as indicated on the top of the panels. The Pearson's linear correlation coefficients between frequency shifts and MAI (from left to right) are 0.94, 0.95, 0.94.} 
    \label{fig:spatial_high} 
\end{figure}
\begin{figure}
\centering
   \includegraphics[width=0.8\textwidth]{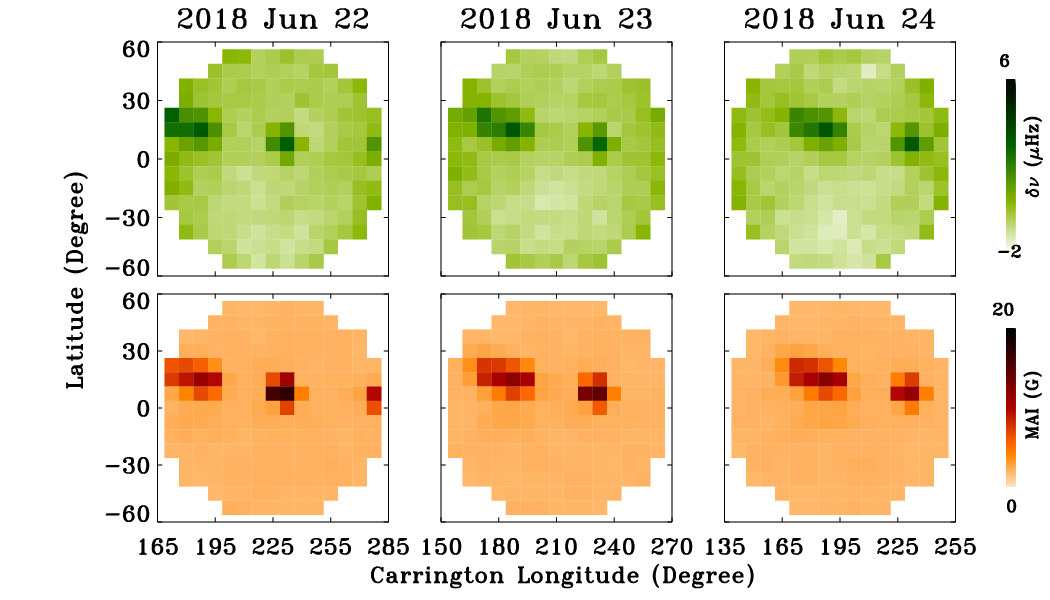}  
   \caption{Spatial distribution of average frequency shifts of 189 dense pack tiles (top) and coeval MAI (bottom) during low activity period in 2018 as indicated on the top of the panels. The Pearson's linear correlation coefficients between frequency shifts and MAI (from left to right) are 0.73, 0.77, 0.78.} 
    \label{fig:spatial_low1}   
\end{figure}

\begin{figure}[t!]
    \centering
\includegraphics[width=0.8\textwidth]{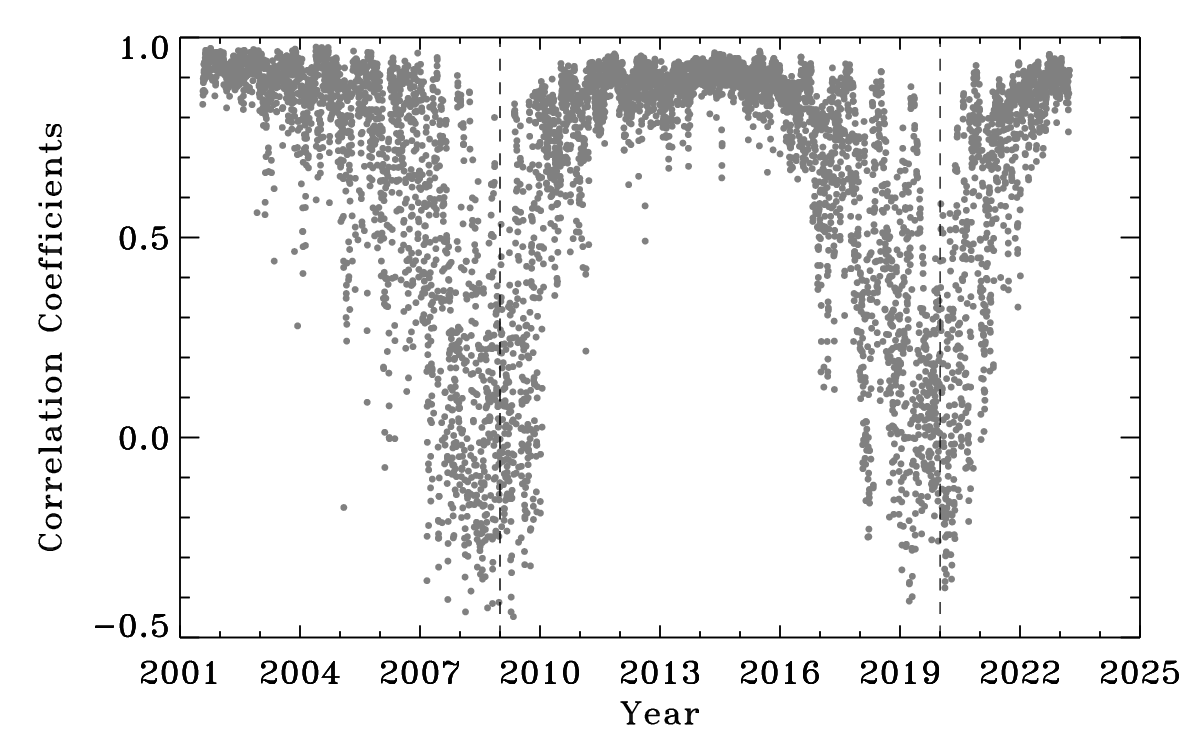} 
\caption{Temporal variation of  Pearson's linear correlation coefficients between  frequency shifts, $\delta\nu_s$, and magnetic activity index, MAI, for each ring-day. The vertical dashed lines
have the same meaning as in Figure~\ref{dutytime}. \label{fig:dailycor}}   
\end{figure}

In Figures~\ref{fig:spatial_high} and ~\ref{fig:spatial_low1}, we display examples of spatial distribution of $\delta\nu$ and the corresponding MAI values for three consecutive days during the high- and low-activity periods in cycle 24, respectively. Each small square in these panels corresponds to a 15$^{\circ}$ $\times$ 15$^{\circ}$ tile. It can be easily seen that the tiles with higher $\delta\nu$ values have good correspondence with those with high MAI values or vice versa. This is quantitatively checked by calculating the Pearson's linear correlation coefficients, $r_p$, and these are found to be 0.94, 0.95 and 0.94 for the high-activity period and 0.73, 0.77 and 0.78 for the low-activity period. The decrease in correlation with low magnetic activity is consistent with the studies mentioned earlier and confirms that the correlation indeed varies during the two different periods of solar activity. 
 Figure~\ref{fig:dailycor} displays the temporal variation of $r_p$ computed between frequency shift and MAI over the dense-pack mosaic of each ring-day. One can easily notice that  $r_p$ values remain high during the high activity period but low during the rising and declining phases of the solar cycle. We also obtain low or negative correlation around the cycle minimum, which is similar to the results obtained with global-mode analysis for the minimum between cycles 23 and 24 \citep{Tripathy10}. 

\begin{figure}[t!]
    \centering
\includegraphics[width=0.8\textwidth]{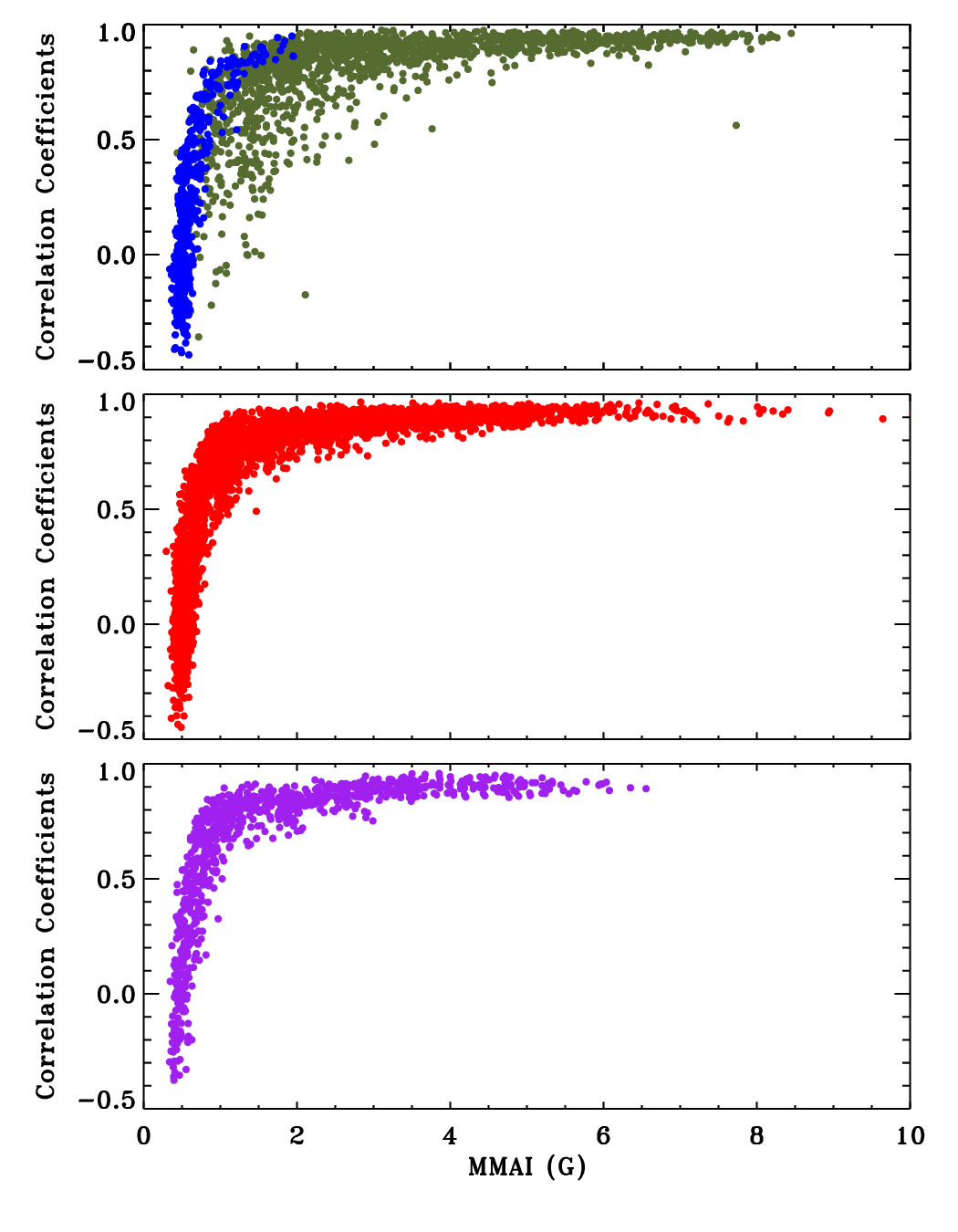}  
\caption{Pearson's linear correlation coefficients, $r_p$, between  $\delta\nu_s$ and mean magnetic activity index, MMAI, for each ring-day as a function of MMAI. Solar cycle 23 values are shown in the top panel; the olive and blue dots depict the MMAI values computed from MDI and GONG magnetograms, respectively.  The middle and bottom panel shows the coefficients for cycles 24 and 25, respectively.  \label{fig:dailycor2}}
\end{figure}

Figure~\ref{fig:dailycor2} illustrates the variation in $r_p$ as a function of MMAI. This reveals that for MMAI higher than 2~G, the correlation coefficients  tend to remain between 0.8 and 1.0.  As substantiated in Paper~I, the correlation coefficients increase with the higher MMAI values and decrease as the solar activity progresses towards the minimum phase characterized by low MMAI values. \citet{Tripathy13} using data for four years near the cycle 24 minimum have also reported a similar pattern in correlation coefficients calculated between the daily values of $\delta\nu_s$ and MMAI.   Based on the argument of weak correlation between low-degree modes and activity proxies, \citet{foullon10} carried out a parametric study and suggested that the anomalous frequency shifts are best explained by thermal effects in the upper regions of the convection zone.  However as outlined in Section~\ref{sec:intro}, the analysis of oscillation frequencies during the solar cycle favors a near-surface magnetic perturbation. Our findings agree with this explanation except around the minimum phase where we observe weak correlation.  We  conjecture that weak  turbulent magnetic fields in the solar photosphere outside of active regions \citep{hanle2001} may be responsible to change the oscillation frequencies during the minimal-activity phase of the cycle. 

As seen in Figure~\ref{fig:dailycor2}, the behaviour is strikingly similar between different solar cycles except for a large scatter observed during solar cycle 23 (top panel). This is due to the use of MDI magnetograms (olive dots) for calculating the MAI values as these magnetograms are derived from a combination of 1-min and 5-min observations with different noise levels.  With the use of a uniform set of merged GONG magnetograms obtained with 32~min cadence beyond 2007 April 09, the scatter in $r_p$ is significantly reduced. We also obtain negative values for $r_p$ when MMAI is below 1~G. To assess  the significance of the negative correlations, we computed the Spearman's rank correlation since it provides the two-sided significance of deviation from zero. It is found that the negative correlation has a two-sided significance value greater than 0.01 which lead us to conclude that these correlations are insignificant.  

\subsection{Temporal Evolution of Mode Frequencies and Activity}\label{sec:evolution}

In order to investigate the temporal variation of mode frequencies, we adopt a method which is analogous to frequency shifts calculated in global $p$-mode analysis \citep{Woodard1991} i.e.
\begin{equation}
    \delta\nu = \Sigma_{n,l}\frac{\Sigma_{i=1}^{189}\nu_{n,l,i}-\overline{\nu_{n,\ell}}}{\sigma_{n,l}^2} \Bigm/ \Sigma_{n,l}\frac{1}{\sigma_{n,l}^2} \;, 
    \label{eq_temp}
\end{equation}
where $\delta\nu$ is the frequency shift over one ring-day and is determined by subtracting the reference frequency, $\overline{\nu_{n,\ell}}$ from the weighted sum of the frequencies of each ($n$, $\ell$) mode averaged over the 189 tiles.    Once again, the weights are chosen as the square of the fitting uncertainties, $\sigma_{n,l}$. It may be noted that while computing the weighted sum of $\nu_{n,l}$ over the dense-pack mosaic, we include  only those modes that are present in at least two-third of the tiles since the fitting procedure does not yield 
the same number of identical modes in each tile or even in each ring-day. Thus, we analyze the oscillation data in degree and frequency range of $200 \le \ell \le 900 $ and $1500 \le \nu \le 5200~\mu$Hz. The mean frequencies as a function of harmonic degree, commonly known as $\ell-\nu$ diagram is shown in appendix~A (Figure~\ref{lnu}).

Although the analysis covers two activity minima, preceding and following cycle 24, the reference frequencies are computed from mode frequencies during the quiet activity period following cycle 24. Since the solar minimum following cycle 24 occurred in December 2019,  the reference frequency of each mode is calculated as an weighted average over the 189 tiles across the period 2019 December 17--31 (12 ring-days).  The associated error, $\sigma_{ref}$, is also determined by propagating the values of the fitted uncertainties over the same interval. As mentioned earlier, identical modes are not fitted in every data set and hence we chose 12 ring-days to compute the reference frequencies to include all possible modes. We further emphasize that 
the results presented below are independent of the choice of reference values since the magnitude of the frequency shifts over the entire data do not change with a different set of reference frequencies. 

The calculation of the errors associated with the frequency shifts consists of two steps. In the first step,  we compute the mean error over the 189 tiles by  using the standard formula for the error of the weighted mean ($\sigma^2 = {1}/{(\Sigma_{n,l}\; {1}/{\sigma_{n,l}^2})}$. In the next step, this mean error and the reference error, $\sigma_{ref}$ are summed through the error propagation formula to yield the formal uncertainty corresponding to $\delta\nu$ of each ring-day.  

Although earlier studies have shown that the observational gaps do not significantly bias the identification of mode frequencies \citep{chaplin04, jeb2019}, for the sake of completeness, we analyze the frequency shifts as a function of the duty cycle. The results reveal no relation between the frequency shifts and duty cycle. We expand on this in Appendix~A.   

Figure~\ref{fig:dnu} shows the frequency shifts (dots) and associated errors as a function of time. Since the magnitude of the errors are too small to be visible in the figure, it is multiplied by a factor of 5 before plotting.  Also to avoid cluttering, errors are drawn for every 250th point.  The dots in the figure reveals a large scatter since we are plotting the daily variation of $\delta\nu$ as opposed to averages over longer time period. Similar scatter is also seen in activity proxies plotted later.  In order to decrease the scatter and visualize different features, we smooth the data through a boxcar averaging over 181 points. The smooth profile shown by the thick black line in the figure  clearly displays  different phases of the solar cycles (the lines in relevant figures throughout the rest of the paper will represent the boxcar averaged smooth profiles).   The other key features include a double-peak structure at the activity maximum of cycle 24,  a minimum following cycle 24 as deep as the minimum following cycle 23, and a smaller peak amplitude in cycle 24 compared to cycle 23. 

\begin{figure}[ht!]
    \centering
\includegraphics[width=0.8\textwidth]{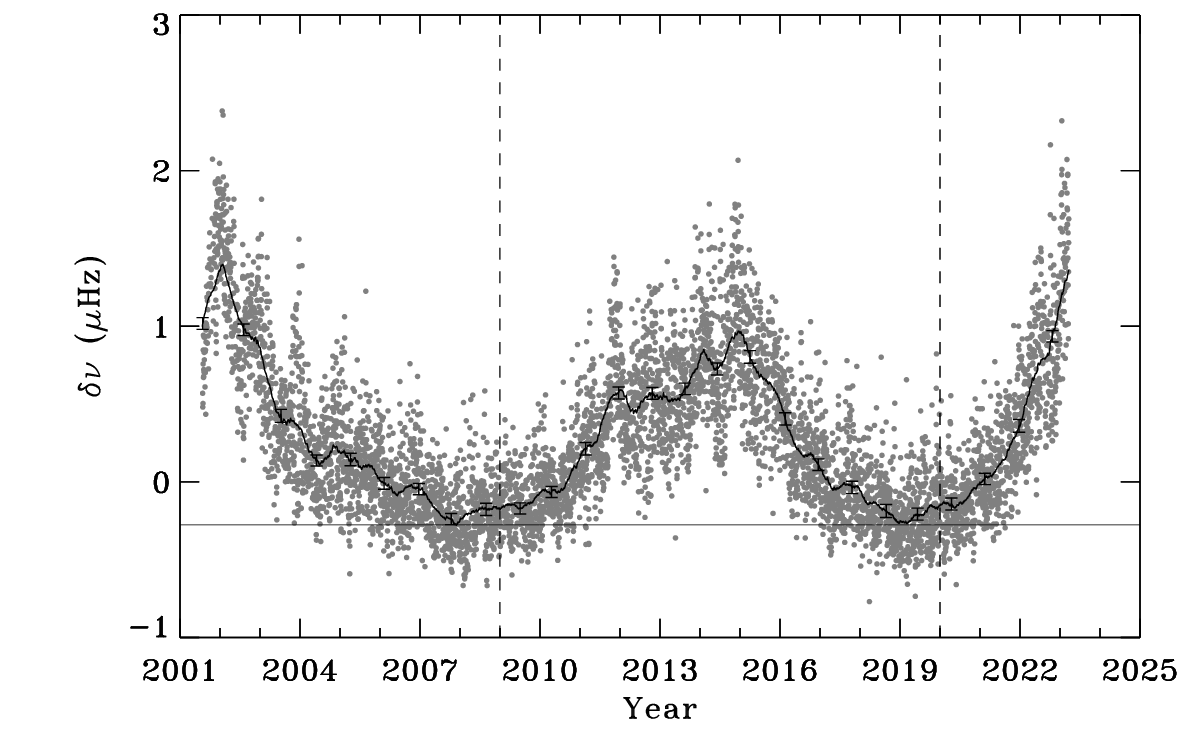}    
\caption{Dots denote the frequency shifts corresponding to each ring-day. The thick black line represents the smoothed profile obtained through a boxcar average of 181 data points.  The horizontal thin line indicates the position of the solar minimum preceding cycle 24. The vertical dashed lines have the same meaning as in Figure~\ref{dutytime}.  To avoid cluttering, errors are shown for every 250th point on the smoothed curve and are  multiplied by a factor of 5 to be visible on this plot. \label{fig:dnu}}
\end{figure}

\begin{figure}[ht!]
    \centering
\includegraphics[width=0.8\textwidth]{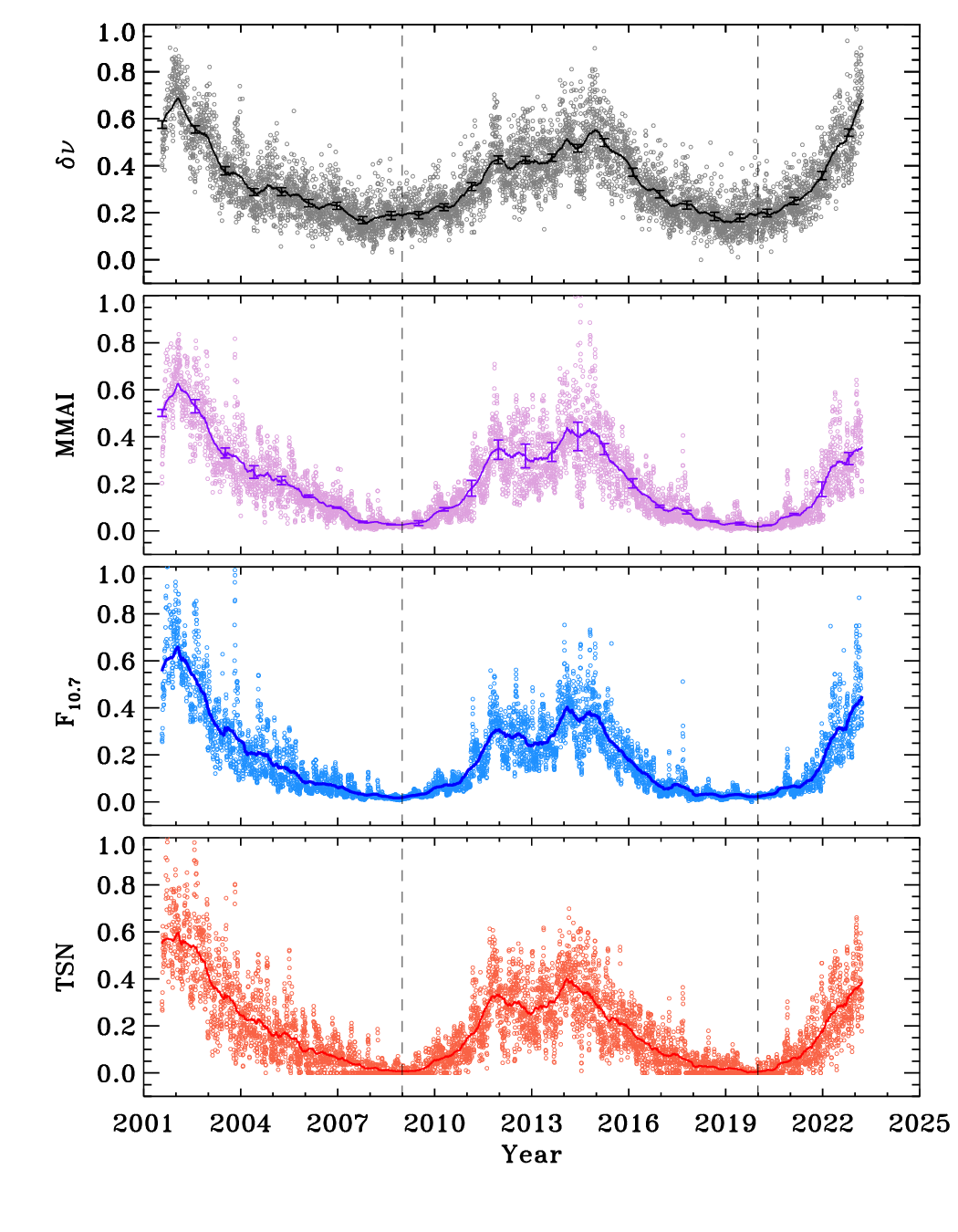}    
\caption{The dots depict the  temporal evolution of frequency shifts, $\delta\nu$, mean magnetic activity index, MMAI, total sunspot number, TSN, and 10.7 cm radio flux, $F_{10.7}$. All quantities are scaled to have values between 0 and 1. The colored lines are the smoothed profiles where smoothing is performed by a boxcar average of 181 data points. The dashed vertical lines
have the same meaning as in Figure~\ref{dutytime}.  To avoid cluttering, errors are shown for every 250th point on the smoothed curve and are  multiplied by a factor of 5 to be visible on this plot. For a plot depicting only the smoothed curves, the reader is referred to Figure~\ref{fig:anuact} in Appendix~A. \label{fig:nuact}}
\end{figure}
Figure~\ref{fig:nuact} exhibits the temporal evolution of $\delta\nu$, MMAI, $F_{10.7}$ and TSN where all quantities are scaled to have values between 0 and 1.  
At a first glance, it is observed  that the activity indices follow the same cyclic trend as frequency shifts. The smooth profiles also illustrate the discrepancy as to the timing of the solar minimum; minimum occurring earlier in frequency shifts compared to the solar activity proxies.   However, the timings of solar maxima appear to coincide. On closer examination, we note that the minimum of cycle 24 seen in activity indices was deeper compared to cycle 25 minimum while, as discussed earlier, the minimum seen in frequency shifts is comparable during both the cycles.   

A similar discrepancy was already noticed during the minimum of solar cycle 24 (paper I) where it was pointed out that the minimum period observed in frequencies persisted for a short period compared to activity indices. In this analysis, we find that during the cycle 25 an early minimum is detected in the frequencies with a lead time of about one year as compared to the activity indices.  This provides added evidence to the interpretation of paper I  that the large changes in high-degree frequencies are caused by the strong field located in the base of the convection zone  while the small-scale,  weak  turbulent magnetic field is responsible for the smaller shifts which occur mostly during the minimum phases of the solar cycle. 
\begin{figure}
\centering
 \includegraphics[width=0.8\textwidth]{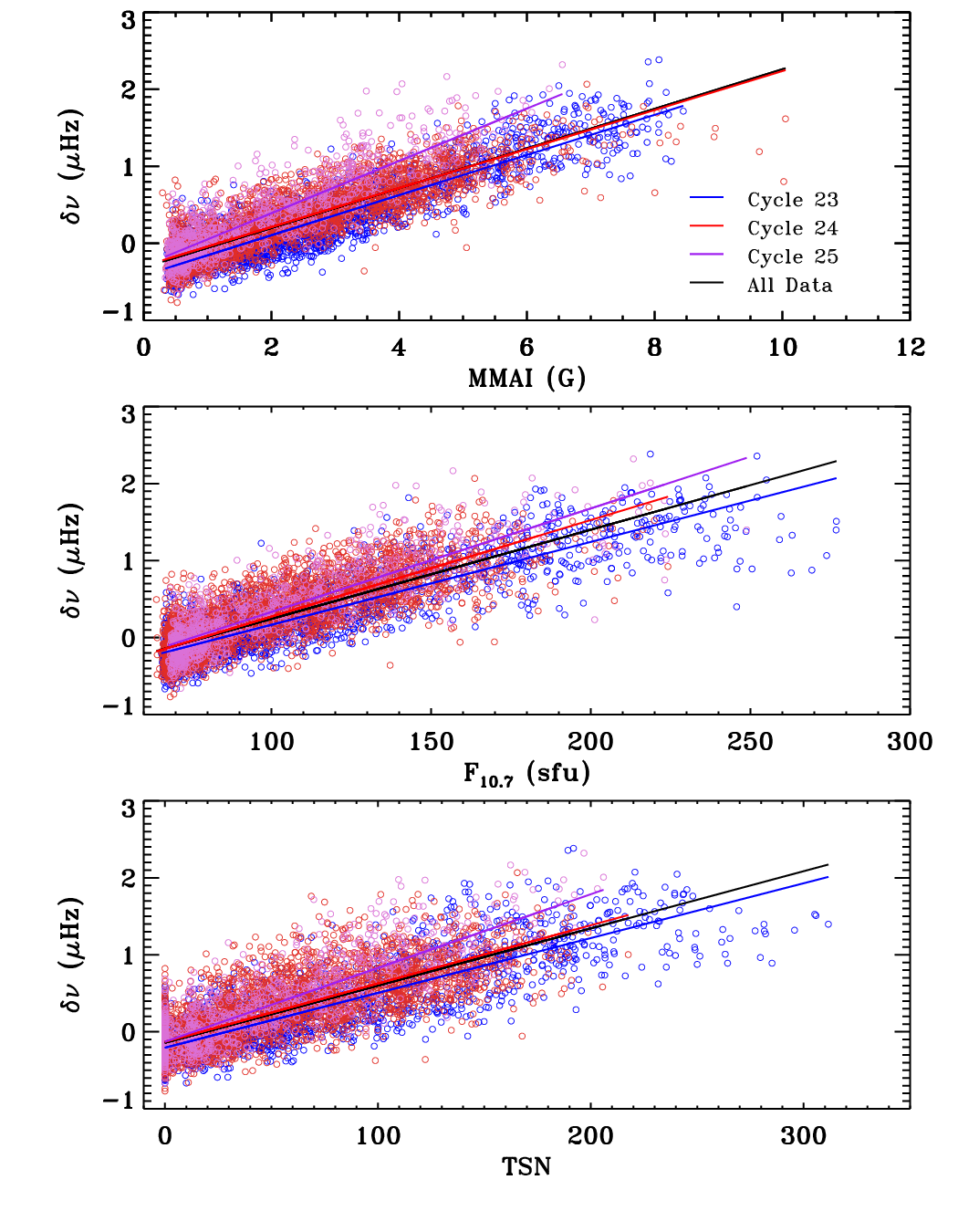}    
\caption{The open circles shows the frequency shifts, $\delta\nu$, as a function of total sunspot number, TSN  (bottom panel),  10.7 cm radio flux, F$_{10.7}$ (middle panel), and mean magnetic activity index, MMAI (upper panel). Different colors represent different cycles and are indicated in the top panel.  
The corresponding colored lines denote fits obtained from the linear regressions. We do not plot the error bars since these are small but their values are tabulated in Table~\ref{T-tab1}.  Note that the red and black lines in the upper panel are mostly overlapping. For a plot illustrating only the linear fits, please see Figure \ref{fig:aslopes} in Appendix~A. 
    \label{fig:slopes}}
\end{figure}

To study the response of the frequency shifts to changing levels of solar activity, we perform linear regression analyses of the frequency shifts with all the three activity indicators.  The results of regression are shown in Figure~\ref{fig:slopes} as a black solid line. We find that the gradients are positive; the smallest and largest values are  obtained for TSN and MMAI, respectively.  For investigating the differences between solar cycles, we perform regressions for individual solar cycles and  the fits are shown in  Figure~\ref{fig:slopes} by different colored lines.  The coefficients of the linear regression  
and the Pearson's linear correlation coefficients are provided in Table~\ref{T-tab1} (We did not calculate the coefficients for solar cycle 25 since the data spans only for about three years). 
Comparing the slopes between different cycles, we further notice that the MMAI has the largest gradient in cycle 23 while for the other two indices, the gradients are larger for cycle 24.  This is not surprising since the MMAI represents the local magnetic activity over the same portion of the solar disk as the frequency shifts while the other two indices represent the activity levels over the entire disk.  Also it may be noted that we do not have data for the whole cycle 23 and hence it is hard to say if this steep slope is inherent to the last half of the solar cycle or unique to all of solar cycle 23. 

As with most past studies, we find that the correlation coefficients with different activity indices is greater than 0.75 which increases to 0.95 or higher when the data was smoothed by a boxcar average of 181 data points.  As anticipated, the correlation is highest between the frequency shifts and magnetic activity measured locally.  We also notice some differences between the cycles, the correlation decreasing in cycle 24 compared to cycle 23;   a decreases of  3.6\% for MMAI, 6.8\% for radio flux and 
8.2\% for TSN. The small decrease in linear regressions and correlation coefficients between cycles 23 and 24  may be a consequence of not having complete data for solar cycle 23.  It will be of interest to carry out a statistical analysis of cycles 24 and 25 when data is available for complete cycle 25.

\begin{deluxetable*}{llcclcclc}
\tabletypesize{\footnotesize} 
\tablewidth{0pt}
\tablecaption{Gradients and Linear Correlation Coefficients between Frequency Shifts and Activity Indices  \label{T-tab1}}
\tablehead{
\colhead{}&\multicolumn{2}{c}{Cycle 23} &\colhead{}&\multicolumn{2}{c}{Cycle 24} &\colhead{}&\multicolumn{2}{c}{All Data}\\
\cline{2-3} \cline{5-6} \cline{8-9}\\
 \colhead{Activity Index}& \colhead{Gradient}&\colhead{$r_p$}&\colhead{}& \colhead{Gradient}&\colhead{$r_p$}
&\colhead{}&\colhead{Gradient}&\colhead{$r_p$}
}
\startdata
MMAI&0.2606 $\pm$ 2.63$\times 10^{-3}$ &0.91 (0.99) &&0.2532 $\pm$ 2.46$\times 10^{-3}$&0.87 (0.98) &&0.2576 $\pm$ 1.86$\times 10^{-3}$&0.87 (0.95)\\
F$_{10.7}$ &0.0108 $\pm$ 1.26$\times 10^{-4}$&0.88 (0.99) &&0.0125 $\pm$ 1.51$\times 10^{-4}$&0.82 (0.97)&&0.0115 $\pm$ 9.40$\times 10^{-5}$&0.84 (0.97)\\
TSN&0.0071 $\pm$ 1.00$\times 10^{-4}$&0.85 (0.99)&& 0.0076 $\pm$ 1.07$\times 10^{-4}$&0.78 (0.95)&&0.0074 $\pm$ 6.90$\times 10^{-5}$& 0.80 (0.95)\\
 \enddata
 \tablecomments{Units of the gradients are $\mu$Hz per unit activity. The numbers in the parenthesis denote the correlation coefficients corresponding to the smooth profiles.} 
\end{deluxetable*} 

\begin{figure}[ht!]
    \centering
\includegraphics[width=\textwidth]{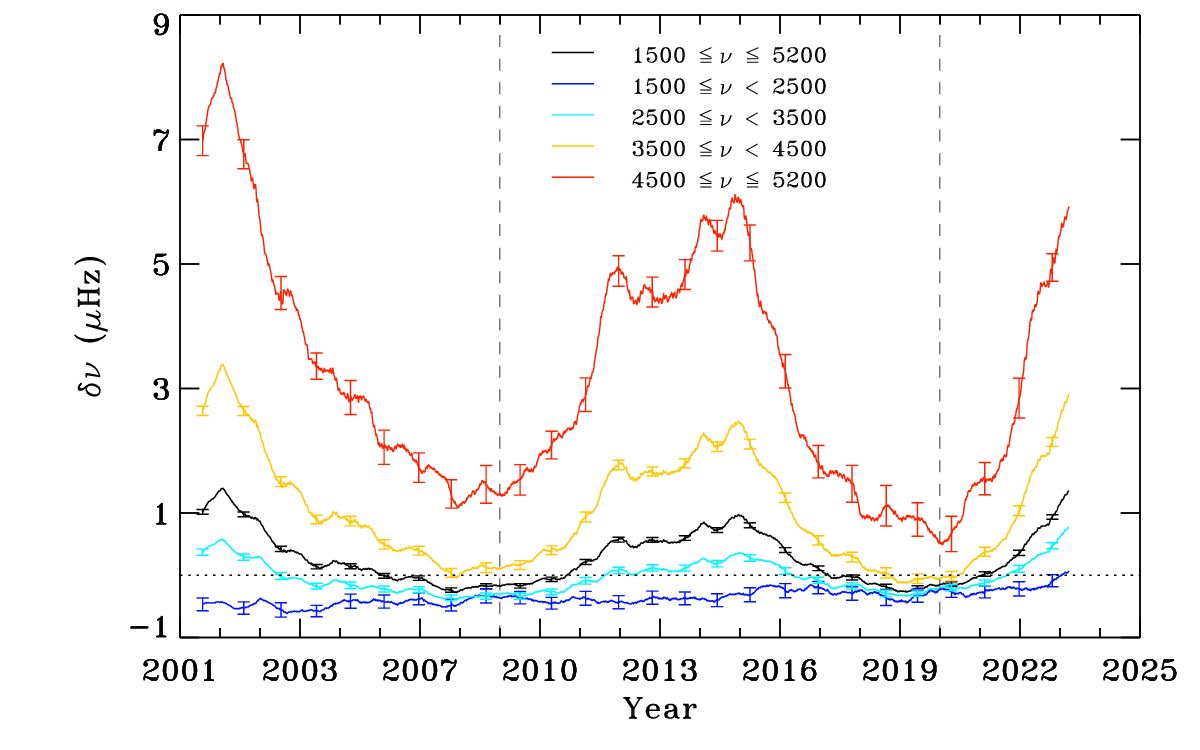}    
\caption{Smoothed frequency shifts, $\delta\nu$, as a function of time for different frequency ranges expressed in $\mu$Hz.  The smoothing is performed with a boxcar average of 181 data points.  
The errors are shown for every 250th point and are  multiplied by a factor of 5 to be visible on this plot.
The dashed vertical lines have the same meaning as in Figure~\ref{dutytime}.  \label{fig:nurange}}
\end{figure}

\begin{figure}[ht!]
    \centering
\includegraphics[width=\textwidth]{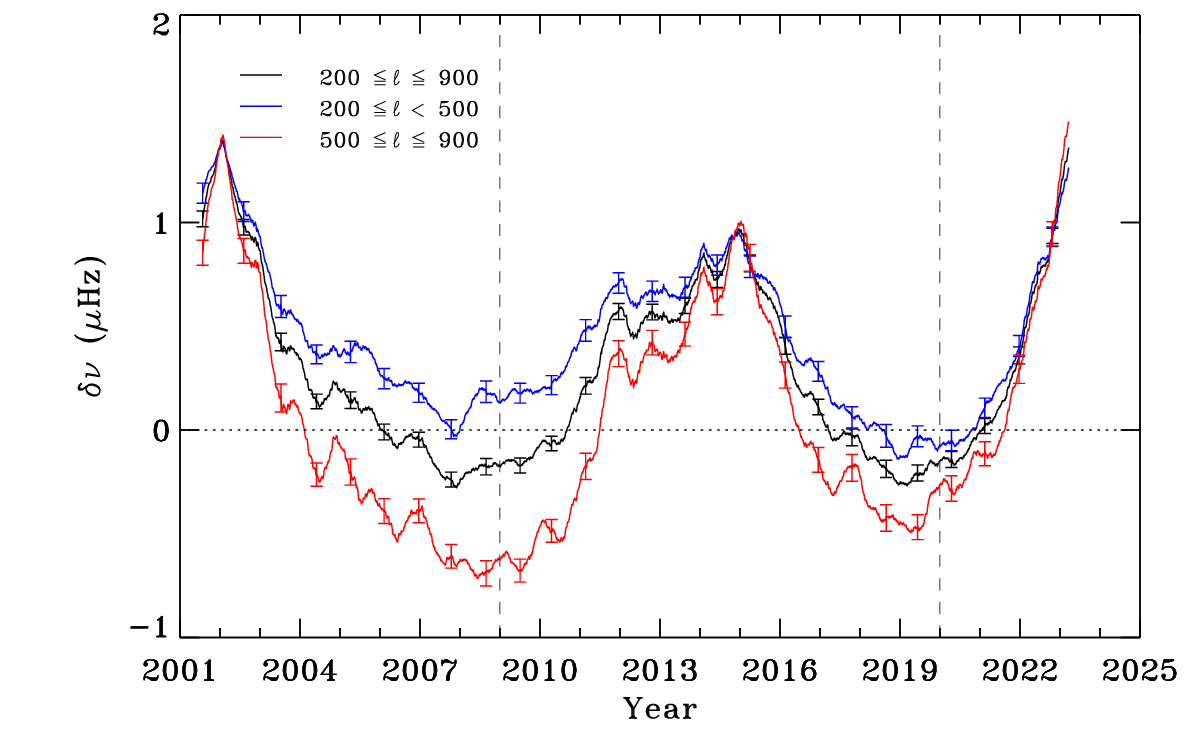}    
\caption{Same as Figure~\ref{fig:nurange}, but for different ranges of spherical harmonic degrees. \label{fig:lrange}}
\end{figure}

\subsection{Frequency Shifts as a Function of Frequency and Degree}
The analysis of global mode frequencies have shown that solar cycle dependent frequency changes 
are predominantly a function of frequency, with the magnitude of shifts increasing with higher frequencies \citep{jain00, broomhall2017} but there are mixed inferences for the degree dependence.  Earlier studies have suggested small or no degree dependence if mode-inertia is taken into account \citep{sct-howe02, salabert04} which implied that solar cycle related changes are confined to a thin layer near the surface. But recent studies involving low-degree modes during the deep minimum period of cycle 24 indicated degree dependence. Particularly, $\ell$ = 0 and 2 modes indicated that the onset of cycle 24 occurred earlier in 2007 while $\ell$ = 1 modes showed the onset to be in early 2009 \citep{sct-salabert09, sct-2010apj, sct11}.   

To determine the frequency and degree dependence of high-degree modes, we calculate the average frequency shifts in four different frequency ranges  with an interval of 1000~$\mu$Hz in $\nu$  and two different degree ranges. The $\nu$ and $\ell$ ranges  are listed in Figures~\ref{fig:nurange} and \ref{fig:lrange}, respectively where we plot the temporal variation of the smoothed frequency shifts. For consistency, the shifts for all ranges are calculated with respect to the same set of reference frequencies corresponding to the 12 ring days during the minimum period of solar cycle 24 (see 
Section~\ref{sec:evolution}).  It is evident that the magnitude of the shifts increase with $\nu$ and $\ell$. To quantify,  we  calculate the difference between the minimum and maximum phase of cycle 24.  These values are shown in Table~\ref{tab2} and  confirm  that the shifts increase with higher frequencies and degree.  Our results are in qualitative agreement with \citet{broomhall2017} where frequency shifts between cycle 23  and 24 minima periods were compared as a function of mode frequency. The figures also reveal  negative shifts in some $\nu$ and $\ell$ ranges implying that the frequencies are lower than the reference frequencies. Thus the choice of using the reference frequencies corresponding to the activity minimum of  solar cycle 24 rendered negative frequency shifts in the two lowest $\nu$ bands and the highest $\ell$  band. Such dependencies of frequency shifts on $\nu$ and $\ell$ have also been observed in global {\it p}-mode analysis and were attributed to the location of upper tuning points from where the modes reflect back into the interior \citep{sct-basu12, Jain22}.   

The frequency shifts also demonstrate a clear solar cycle pattern except for the lowest frequency range of 1500 -- 2500 $\mu$Hz. \citet{chaplin01} have also reported a flat sensitivity below 2500 $\mu$Hz. Table~\ref{tab2} shows the Pearson's linear  correlation coefficients between frequency shifts and activity proxies. We find that the frequencies are highly correlated with the solar activity proxies except for 
the lowest frequency range where a signature of anti-correlation is  seen. However, given that the coefficients are small, we  interpret this as no correlation. 
\begin{deluxetable*}{crrrr}
\tablewidth{0pt} 
\tablecaption{Observed Differences in Frequency Shifts, $\delta\nu$,  between the Maximum and Minimum Phase of Cycle 24
and  Pearson's Linear Correlation Coefficients, r$_p$, between $\delta\nu$ and Activity Proxies.  \label{tab2}}
\tablehead{
\colhead{}&\colhead{Differences in $\delta\nu$\,(nHz)}&\multicolumn{3}{c}{r$_p$}\\
\cline{3-5}
\colhead{}&\colhead{} &\colhead{MMAI}&\colhead{F$_{10.7}$}&\colhead{TSN}
}
\decimals
\startdata
Frequency range ($\mu$Hz) && &&\\
\cline{1-1}
1500 $-$ 5200&1246.4 $\pm$ \phn8.8&\phn0.95&0.96&0.95\\
1500 $-$ 2500&\phn 462.7 $\pm$ 20.1&-0.19&-0.13&-0.17\\
2500 $-$ 3500&\phn 743.4 $\pm$ \phn9.8&\phn0.87&0.90&0.88\\
3500 $-$ 4500&2514.2 $\pm$ 16.1 &\phn0.98&09.8&0.97\\
4500 $-$ 5200&5028.3 $\pm$ 69.2 &\phn0.99&0.98&0.98\\
\hline
Degree range & & &\\
\cline{1-1}
\phn200 $-$\phn900&1246.4 $\pm$ \phn8.8&\phn0.95&0.96&0.95\\
\phn200 $-$\phn500&986.9 $\pm$ \phn9.4 &\phn0.96&0.96&0.95\\
\phn500 $-$\phn900&1719.4 $\pm$ 12.2 &\phn0.90&0.93&0.90\\
\enddata 
\tablecomments{Differences in $\delta\nu$ and $r_p$ are based on the smoothed  values obtained from the boxcar average of 181 data points.} 

\end{deluxetable*}
\subsection{Hemispheric Asymmetry}
Hemispheric asymmetry observed in sunspots or other solar activity proxies is an important aspect of solar cycles as it provides important clues for understanding dynamical processes in the interior of the Sun. However, its origin is still not well understood. Observations illustrate that the major changes in hemispheric flux occur around activity maximum in each cycle due to the cancellation or transport of flux across the equator through the meridional flow \citep{petrie2023}. Simulations reveal that hemispheric asymmetry could result due to the presence of an equator-symmetric part in the oscillating magnetic field \citep{mandal19} or due to a stochastic  process in the solar dynamo suggesting that the solar magnetic field has some memory in the   convection zone \citep{das22}.  Contrary to this, \citet{veronig21} argue that the solar cycle evolves independently in the two hemispheres and suggest that the solar cycle prediction methods can be improved by including the evolution of hemispheric sunspot number.   
\begin{figure}
    \centering
  \includegraphics[width=0.9\textwidth]{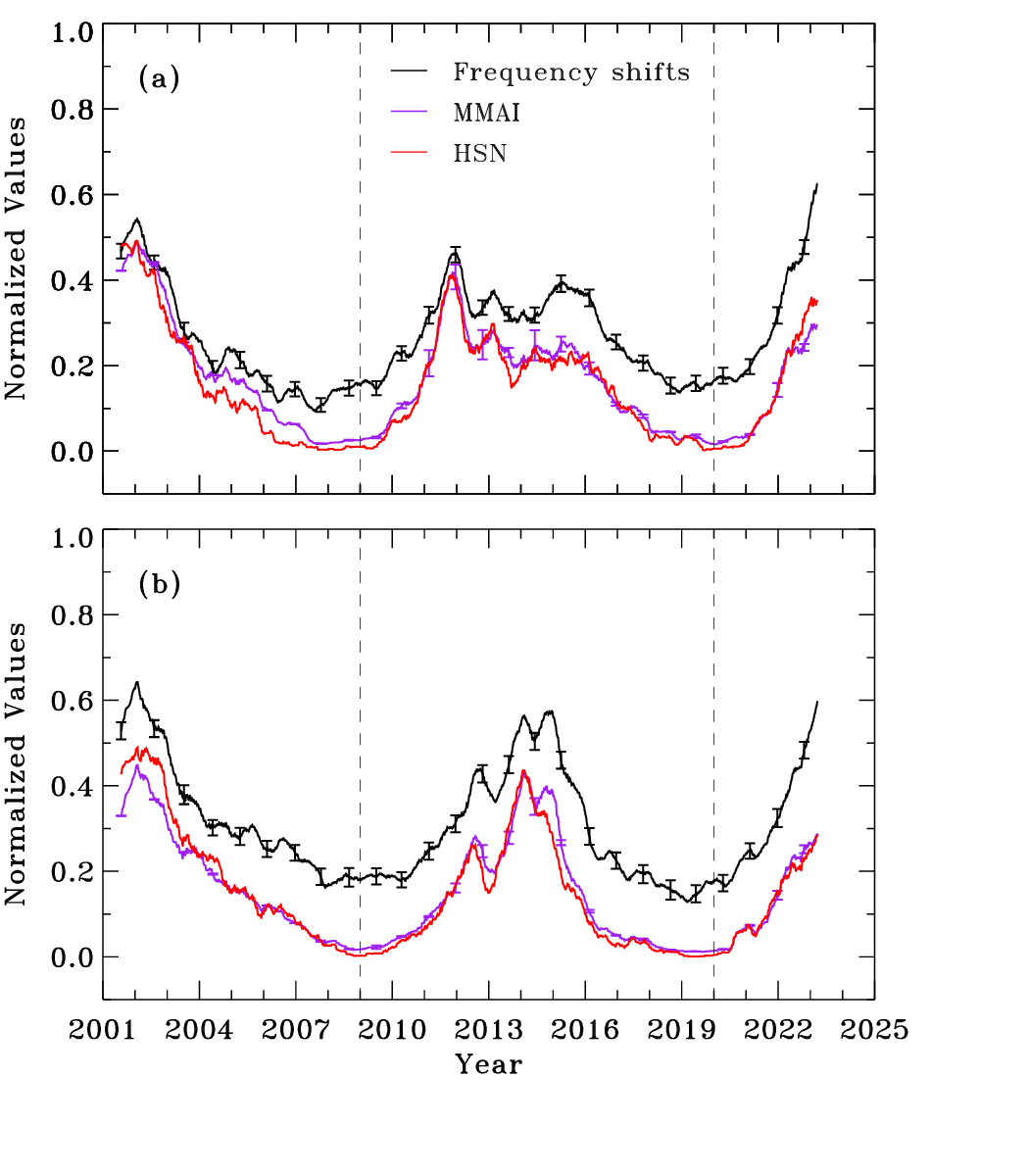}  
    \caption{Temporal evolution of frequency shifts, $\delta\nu$, mean magnetic activity index, MMAI, and hemispheric sunspot number, HSN, over the (a) Northern and (b) Southern Hemispheres. All quantities are scaled to have values between 0 and 1. The colored lines depict the smoothed profiles obtained by applying a boxcar average over 181 data points. 
    The errors for $\delta\nu$ and MMAI are shown for every 250th point and are  multiplied by a factor of 5 to be visible on these plots. The dashed vertical lines have the same meaning as in Figure~\ref{dutytime}.
        \label{fig:hemis}}
\end{figure}

\begin{figure}
    \centering
  \includegraphics[width=0.9\textwidth]{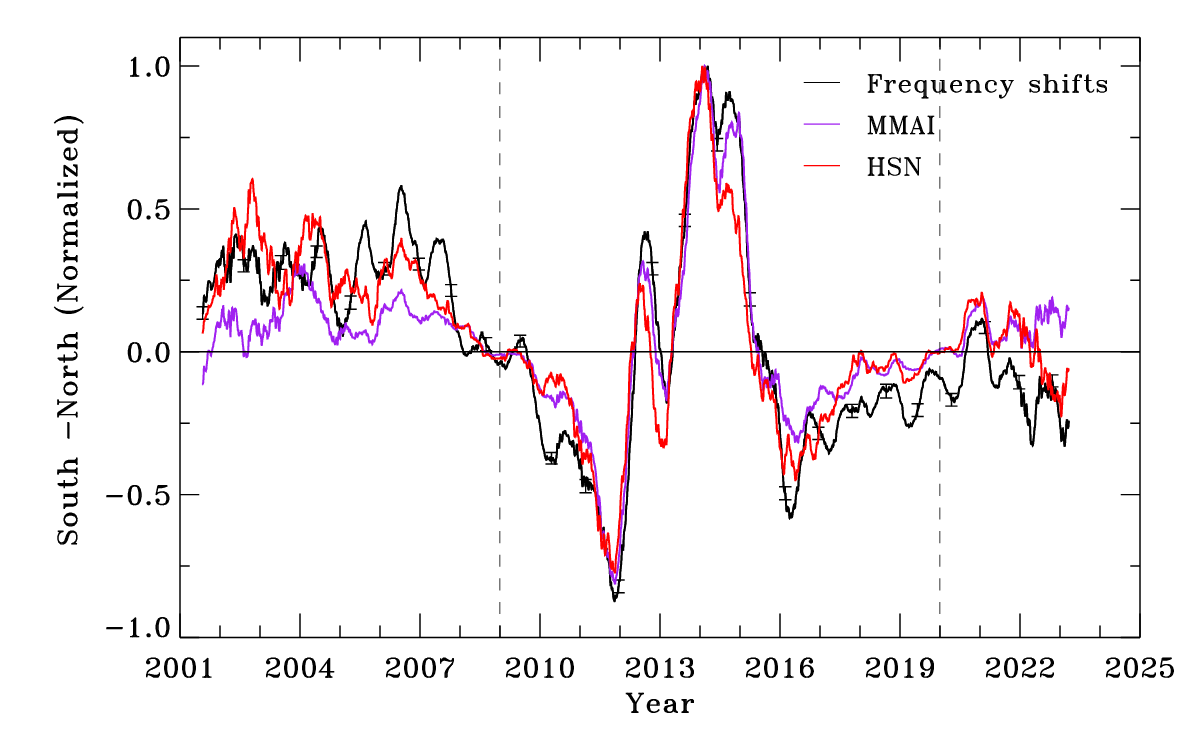}    
    \caption{The difference between south and north components of frequency shifts, $\delta\nu$, mean magnetic activity index, MMAI, and hemispheric sunspot number, HSN, normalized by their maximum values. The colored lines depict the smoothed profiles obtained by applying a boxcar average over 181 data points. 
    The errors for $\delta\nu$ and MMAI are shown for every 250th point and are  multiplied by a factor of 5 to be visible on these plots. The dashed vertical lines have the same meaning as in Figure~\ref{dutytime}.
            \label{fig:n-s}}
\end{figure}

In this context, we investigate the presence of hemispheric asymmetry, especially during  solar cycle 24, by analyzing the frequency variations over the two hemispheres. It may be noted that the {\it m}-averaged central frequencies obtained in global analysis are latitudinally averaged and can not be used to study variations between hemispheres. Figure~\ref{fig:hemis} shows the frequency shifts, MMAI and HSN for both the hemispheres separately, where each quantity is scaled to have values between 0 and 1.  The frequency shifts are calculated using  equation~\ref{eq_temp} where we use the same reference period but averaged  over each  hemisphere separately.  The solid lines representing the smooth profiles illustrate the  asymmetric nature of these quantities.  The frequency shifts closely follow the asymmetry pattern seen in HSN and MMAI, and attain a peak amplitude in late 2011 in the northern hemisphere, two years earlier than the south. We also note two distinct and prominent peaks (double peaks) of similar magnitudes in $\delta\nu$ and MMAI in the southern hemisphere but only one strong peak in the northern hemisphere. The secondary peaks in the northern hemispheres have reduced amplitudes as compared to the main peak. As solar activity decreases with time, the amplitude of frequency shifts is also found to be decreasing in both hemispheres but  rapidly in the southern hemisphere.  
These findings indicate that the magnetic activity in each hemispheres evolve independently and frequency shifts follow this evolution closely. Comparing the total magnetic energy of the Sun between the two hemispheres, \citet{inceoglu17} had also concluded that the two hemispheres are decoupled. Our study supports this assertion.   

For a better understanding of the behaviour of hemispheric asymmetry, we compute the difference between the two hemispheres as south$-$north and present the result in Figure~\ref{fig:n-s}. Here positive  values imply that the  dominant contribution is from the southern hemisphere and negative values from northern hemisphere.  It is evident that during cycle 24,  the frequency shifts vary in phase with MMAI and HSN attaining the maximum (minimum)  amplitude around the same epoch but some small deviations are seen during cycle 23. Nevertheless, the Pearson's correlation coefficient is found to be greater than 0.90 for the entire period confirming a close linear relationship between the shifts and activity proxies in each hemisphere.  Additionally both the  frequency shifts and level of solar activity are found to be dominant in the southern hemisphere during the descending phase of cycle 23 while cycle 24 is characterized by a three-part structure: (i) the northern hemisphere is found to be dominant during the ascending phase of the cycle  (ii) the southern  hemisphere changes to be  dominant during the maximum phase, and (iii) the southern activity declines during the descending phase of the cycle 24 and the northern hemisphere turns to be dominant until the cycle reaches the minimum period. With the start of the new cycle 25, the southern hemisphere is emerging to be dominant in activity proxies and frequency shifts. Remarkably, these results agree well with the study of \citet{petrie2023} where  line-of-sight magnetograms were used to detect the large hemispheric flux fluctuations during cycle 24.  

\begin{figure*}  
   \centering
   \includegraphics[width=0.7\textwidth,angle=90]{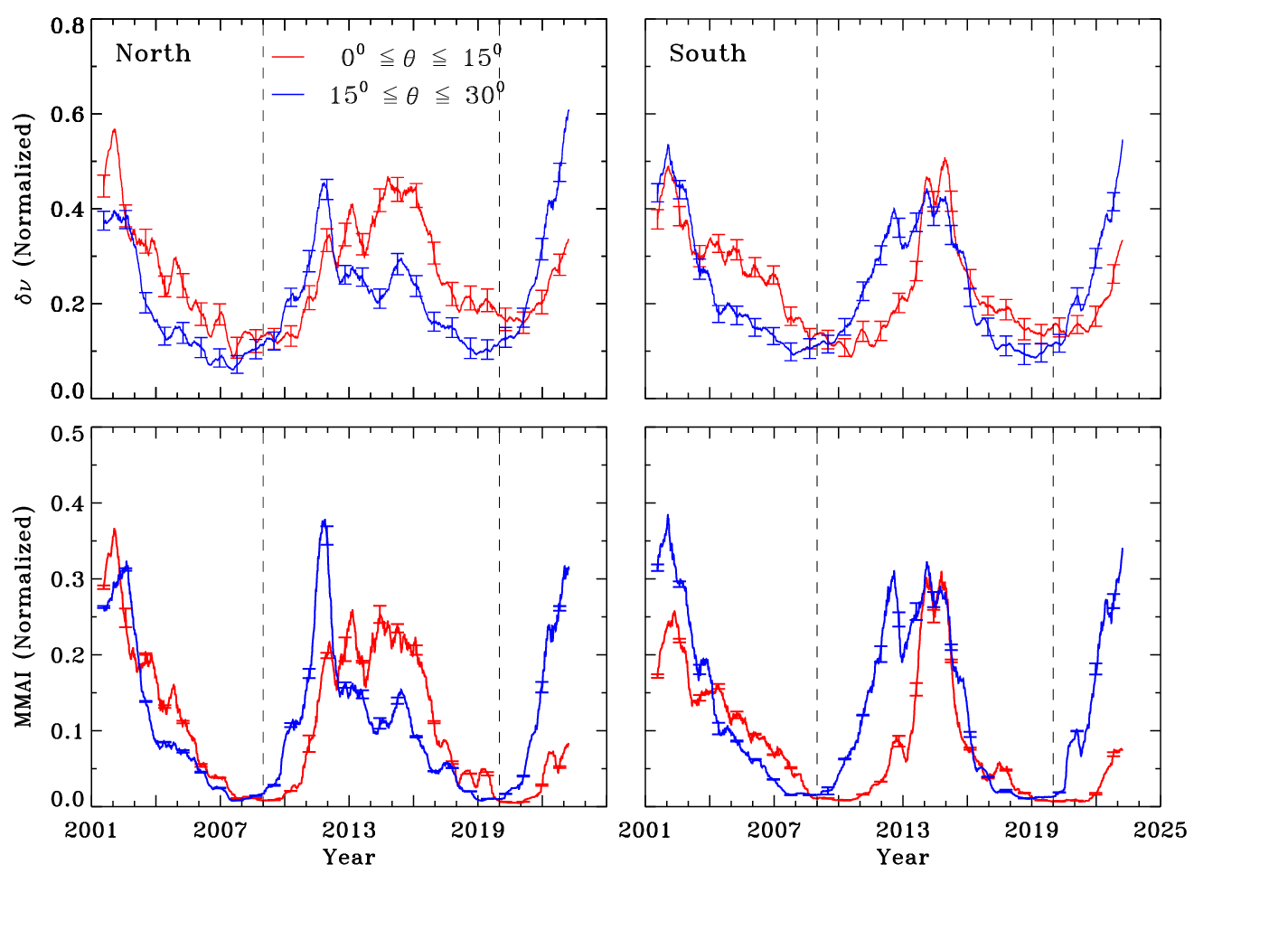}   
   \caption{Top panels display the progression of the solar cycles  as seen in frequency shifts, $\delta\nu$, at two different latitude bands for northern (left) and southern hemispheres (right). The bottom panels illustrate corresponding variations in the mean magnetic activity index, MMAI.  All quantities are scaled to have values between 0 and 1. We have only shown the smoothed profiles obtained by applying a boxcar average over 181 data points. 
    The errors for $\delta\nu$ and MMAI are shown for every 250th point and are  multiplied by a factor of 5 to be visible on these plots. The dashed vertical lines have the same meaning as in Figure~\ref{dutytime}.  }
  \label{fig:lat_mai1}
\end{figure*}
\begin{figure*}   
    \centering
   \includegraphics[width=0.7\textwidth,angle=90]{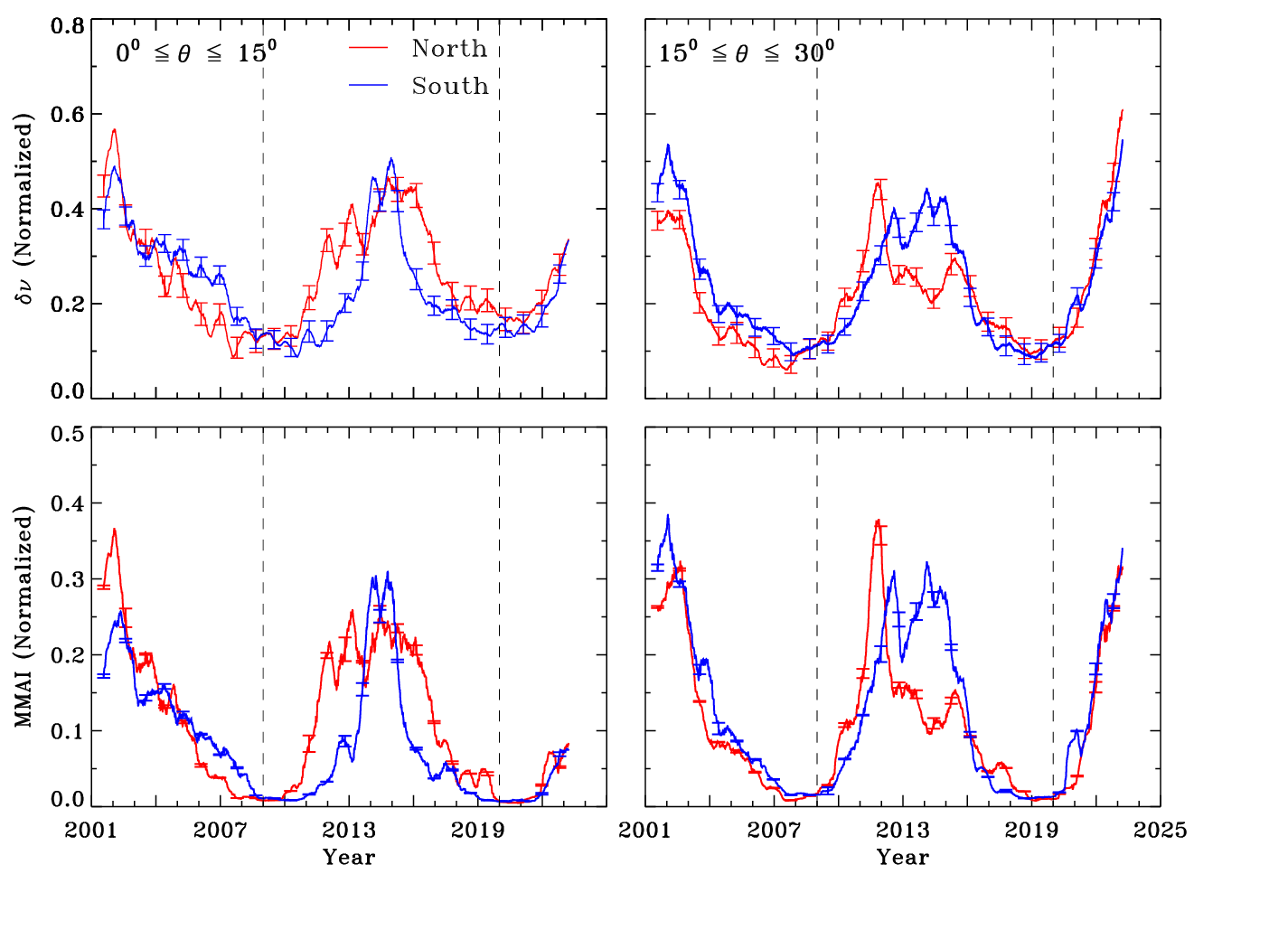}   
    \caption{Top panels display the progression of solar cycles as seen in the smoothed frequency shifts, $\delta\nu$,  in northern and southern hemispheres at two different latitude bands;  $0^\circ \le \theta \le 15^\circ$ (left) and $15^\circ \le \theta \le 30^\circ$ (right). The bottom panels illustrate corresponding variations in the mean magnetic activity index, MMAI. All quantities are scaled to have values between 0 and 1. The smoothed values are obtained  by applying a boxcar average over 181 data points. 
    The errors for $\delta\nu$ and MMAI are shown for every 250th point and are  multiplied by a factor of 5 to be visible on these plots. The dashed vertical lines have the same meaning as in Figure~\ref{dutytime}.    }
    \label{fig:lat_mai2}
\end{figure*}

\subsection{Progression of the Solar Cycle at Different Latitudes} 
We aim at characterizing the frequency shifts during the progression of solar magnetic cycle at different latitudes to explore similarities or differences between them.  Since the cycle starts at mid-latitudes and subsequently migrates equatorward (or poleward), we chose two latitude bands, $0^\circ \le \theta \le 15^\circ$ and $15^\circ \le \theta \le 30^\circ$ (we will refer them as low- and mid-latitude bands hereafter)  as representative latitudes to cover both the start of the cycle and its progress towards the equator. Figure~\ref{fig:lat_mai1} displays the temporal evolution of the frequency shifts (upper panels) and MMAI (bottom panels) in the northern and southern hemispheres for these latitudinal bands separately. While in the northern hemisphere the frequency shifts are almost comparable at the two selected latitudes (the level of activity is marginally stronger), the shifts in the southern hemisphere are higher at the maximum phase in the mid-latitude band but reach similar values in both the bands around  2003 December. Soon after the extended minimum period, the frequency shifts between the two latitudes start to diverge and follow different patterns. The start of the solar cycle 24 is evident in the mid latitude band where the frequencies increased rapidly compared to values in the low latitude band. This reveals a considerable delay in the onset of the cycle in both hemispheres at the lower latitudes.  \citet{simo16} found similar results using both global and high-degree mode frequencies. With the progression of the solar cycle towards the maximum phase, we do not find any  anomaly as the frequency shifts reach their maximum values at mid latitudes first. However, we do observe a delay in the peak values between the two hemispheres probably due to the hemispheric asymmetry seen in the magnetic activity.  Additionally, the shifts corresponding to the low latitude band also demonstrate delay between the two hemispheres; the peak values are reached about a year earlier in the northern hemisphere.  With the onset of cycle 25, as expected, the frequency shifts increased at the mid latitudes in both hemispheres while the shifts remained low at the lower latitudes manifesting overlapping periods between the two cycles.  
Although some differences are observed between the temporal evolution of MMAI and $\delta\nu$, the general behavior between them remains consistent throughout the multiple cycles with a Pearson's linear correlation coefficient value greater than 0.95.  

When we compare the strength of the frequency shifts at the same latitude but for different hemispheres (Figure~\ref{fig:lat_mai2}), we notice that during the descending phase of cycle 23 the southern hemisphere (shown in blue) shifts are higher than the northern ones (red curve). This traces the strength of the magnetic activity. We further note differences in the timing of the minimum phase between the hemispheres as well as low- and mid-latitudes \citep[also see, Paper I and ][]{simo16}. During cycle 24 when the activity started at higher latitudes, the shifts in the southern hemisphere are higher compared to the northern ones while the MMAI  were comparable in both hemispheres. This corroborates that the shifts and magnetic activity  are poorly correlated during low activity periods. But soon after 2010, the frequency shifts in the mid latitude band of the northern hemisphere increase rapidly and peaked in 2011 November  while the shifts in the southern hemisphere peaked during 2014 February. Thus there is a delay of more than 3 years between the two peaks. However, the track of the frequency shifts in the low latitude band is different; the shifts are higher in the northern hemisphere during the ascending phase until 2013 August. After this epoch, the frequency shifts of the southern hemisphere became dominant and reached its peak value in 2014 December   while the peak value in the northern hemisphere was reached around 2014 October.  During the maximum period of the cycle 24 (2014 October-December), the activity levels and the frequency shifts in the mid latitude branch of the southern hemisphere increased and continued to be dominant until 2016 February. As the activity levels further reduced, the frequency shifts grouped together and remained in that configuration till the minimum was reached. A similar overlapping pattern is also seen in the low latitude branch starting from 2018 March. Form the figure it is also evident that the low- and mid-latitude frequency shifts reached the minimum of cycle 25 in two different epochs. Table~\ref{T-tab3} lists the timing of the minimum at these two latitude bands in both southern and northern hemispheres. 
Comparing the timing of the minimum between the two hemispheres we find that there is a delay of about 4 to 6 months for both the frequency shifts and magnetic activity which is shorter compared to the delay observed in cycle 23 as outlined in Paper I. The only difference  is at the mid latitudes where the frequencies of the northern hemisphere reached the minimum phase earlier than the southern hemispheres with a lead time of about 3 months.   
Finally we note that at the beginning of cycle 25 both the frequency shifts and magnetic activity are found to be dominant in the southern hemisphere.
\begin{deluxetable*}{ccccccc}
\tablewidth{0pt} 
\tablecaption{Epochs of Minimum between Cycles 24 and 25 as seen in $\delta\nu$ and MMAI in the Two Hemispheres at Low- and Mid-Latitude Bands and Differences between the Epochs.  \label{T-tab3}}
\tablehead{
\colhead{Latitude }&\multicolumn{2}{c}{$\delta\nu$} &\colhead{ Difference}&\multicolumn{2}{c}{MMAI}&\colhead{Difference}\\
\cline{2-3} \cline{5-6}
\colhead{}& \colhead{South}&\colhead{North}&\colhead{(months)}&\colhead{South}&\colhead{North}&\colhead{(months)}
}
\startdata
$0^\circ \le \theta \le 15^\circ$&08/2020&02/2021&6&06/2020&11/2020&5\phn\\
$15^\circ \le \theta \le 30^\circ$&03/2019&12/2018&-3&11/2018&03/2019&4\phn\\
\enddata
\tablecomments{Positive and negative differences indicate lead and lag time between the two hemispheres.}
\end{deluxetable*}
\section{Summary}
Using  high-degree acoustic mode frequencies from the Global Oscillation Network Group, we investigated how the frequencies,  which carry information from the sub-surface layers, connect to the observed surface activity.   The data span over 22 years starting from 2001 July and include the maximum and descending phase of cycle 23, cycle 24 and three initial years of cycle 25.  We find a strong correlation between the  change in frequencies and the measures of surface activity except during the minimum phase when the local magnetic activity level decreases below 2~G. The correlation remains significantly high on both spatial and temporal scales. 

 Investigating the frequency shifts as  a function of $\nu$ and $\ell$, we find that the magnitude of the shifts increase with $\nu$ and $\ell$ and are in qualitative agreement with \citet{broomhall2017} where the analysis was carried out for global modes. We have also analyzed the shifts as a function of hemispheres separately, and observe 
that the shifts are dominant in southern hemisphere during the  descending phase of cycle 23,  while  a three-part structure is seen in cycle 24; the shifts are dominant in the  northern hemisphere in the initial phase of the cycle, flips to  the southern hemisphere during the maximum period, and to the northern hemisphere during the descending phase of the cycle 24. For the initial period of cycle 25, we detect the frequency shifts in southern hemisphere to be dominant. Since the hemispheric frequency shifts closely follow the behavior of  hemispheric sunspot number and local magnetic activity index, it demonstrates a strong relationship between the surface activity proxies and the near surface shear layer where the resonant modes have lower-turning points.  Our analysis also highlights the differences in the  progression of solar cycle at two different latitudinal bands. The frequency shifts in mid-latitudinal band are higher than the low-latitudinal bands during the rising phase of the solar cycle, however they become comparable once the maximum phase is reached. This is in agreement with the appearance of sunspots during the cycle as they emerge first at higher latitude and then migrate towards the equator. 
Since the evolution of the frequency shifts observed between the two hemispheres have different characteristics, we concur with a few recent assertions that the magnetic activity in two hemispheres are decoupled and evolve independently. Based on the analyses presented here, we emphasize that the frequency shifts observed in high-degree modes are caused by a combination of strong fields present at the tachocline and weak turbulent fields present in the near surface shear layer.  

\section{Acknowledgments}
We thank the reviewer for several useful comments.  MB thanks Boulder Solar Alliance for providing the REU opportunity (NSF grant 1950911)  and the National Solar Observatory. This work utilises GONG data obtained by the NSO Integrated Synoptic Program, managed by the National Solar Observatory, which is operated by the Association of Universities for Research in Astronomy (AURA), Inc. under a cooperative agreement with the National Science Foundation and with contribution from the National Oceanic and Atmospheric Administration. The GONG network of instruments is hosted by the Big Bear Solar Observatory, High Altitude Observatory, Learmonth Solar Observatory, Udaipur Solar Observatory, Instituto de Astrof\'{\i}sica de Canarias, and Cerro Tololo Interamerican Observatory.  We also use data from the Solar Oscillations Investigation/Michelson Doppler Imager on the Solar and Heliospheric Observatory. SOHO is a mission of international cooperation between ESA and NASA. The sunspot data is courtesy of WDC-SILSO, Royal Observatory of Belgium, Brussels.

\facilities{GONG, SoHO/MDI}
\software 
{Interactive Data Language (IDL)\footnote{\url{https://www.nv5geospatialsoftware.com/docs/home.html}}.}

\appendix

\section{Effect of Observational Gaps on Mode Frequencies}
\begin{figure}[th!]
    \centering
\includegraphics[width=0.8\textwidth]{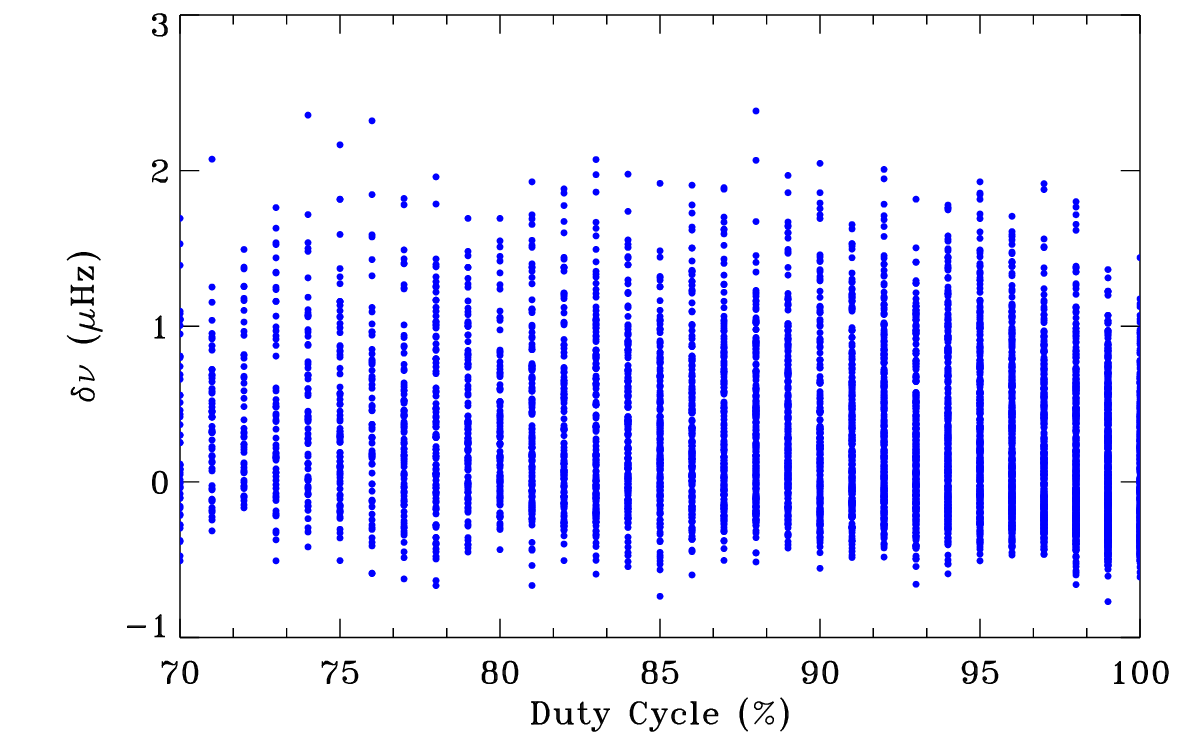}  
\caption{Frequency shifts, $\delta\nu$,  as a function of duty cycle expressed in percentage.} \label{dutynu}
\end{figure}

It is well known that gaps in observation leads to aliasing in the frequency domain and create spurious side lobes that lead to ambiguous identification of real frequencies and other mode parameters. However, analysis of low- and intermediate-degree mode parameters have indicated  
that the bias in estimating the mode frequencies is not significant while it is important 
for mode amplitudes and widths \citep{chaplin04, Broomhall2015, Kiefer18, jeb2019}. However, as a sanity check, we have studied the frequency shifts as a function of the duty cycle and the results are shown in Figure~\ref{dutynu}. As expected, we do not find any relation between frequency shifts and the duty cycle confirming that the duty cycle does not introduce any consequential bias  in the identification of high-degree modes.  
\newpage
\section{supplemental Figures}
\begin{figure}[ht!]
    \centering
\includegraphics[width=0.8\textwidth]{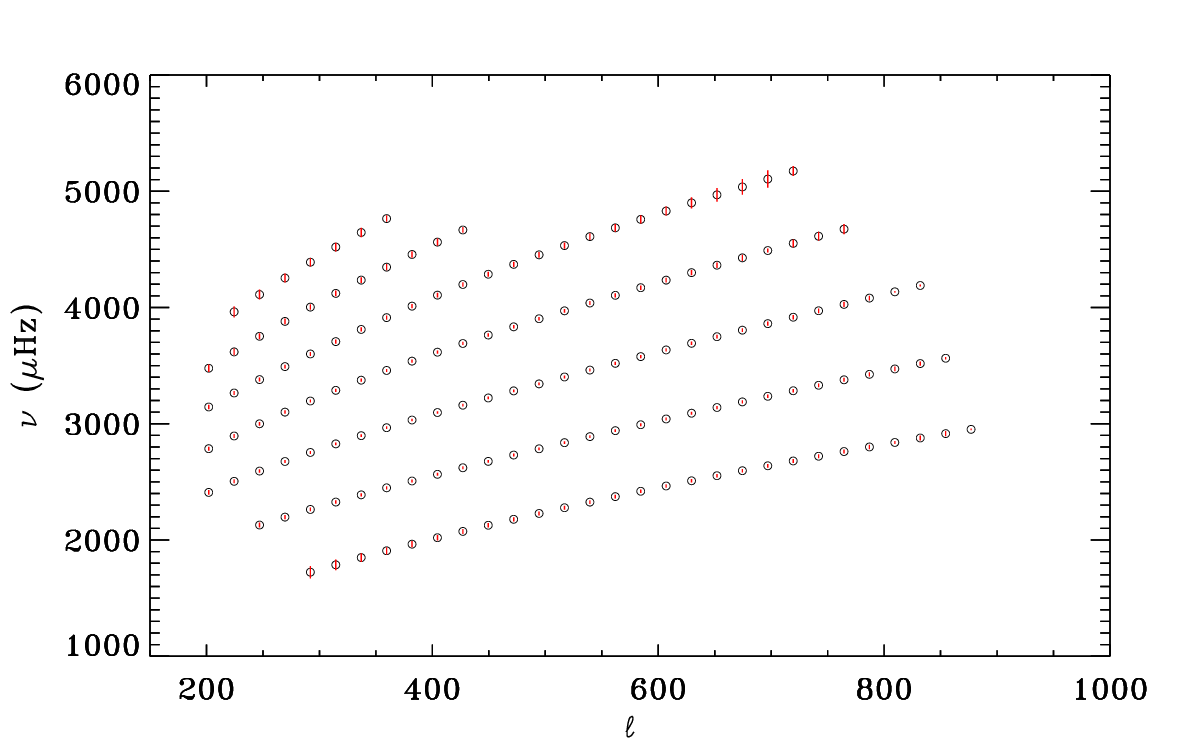}  
\caption{Distribution of fitted modes in $\ell-\nu$ plane (commonly referred as $\ell-\nu$ diagram) corresponding to the reference frequency which is obtained from an average over 12 ring days corresponding to 2019 December 19-31.} \label{lnu}
\end{figure}

\begin{figure}[ht!]
    \centering
\includegraphics[width=0.8\textwidth]{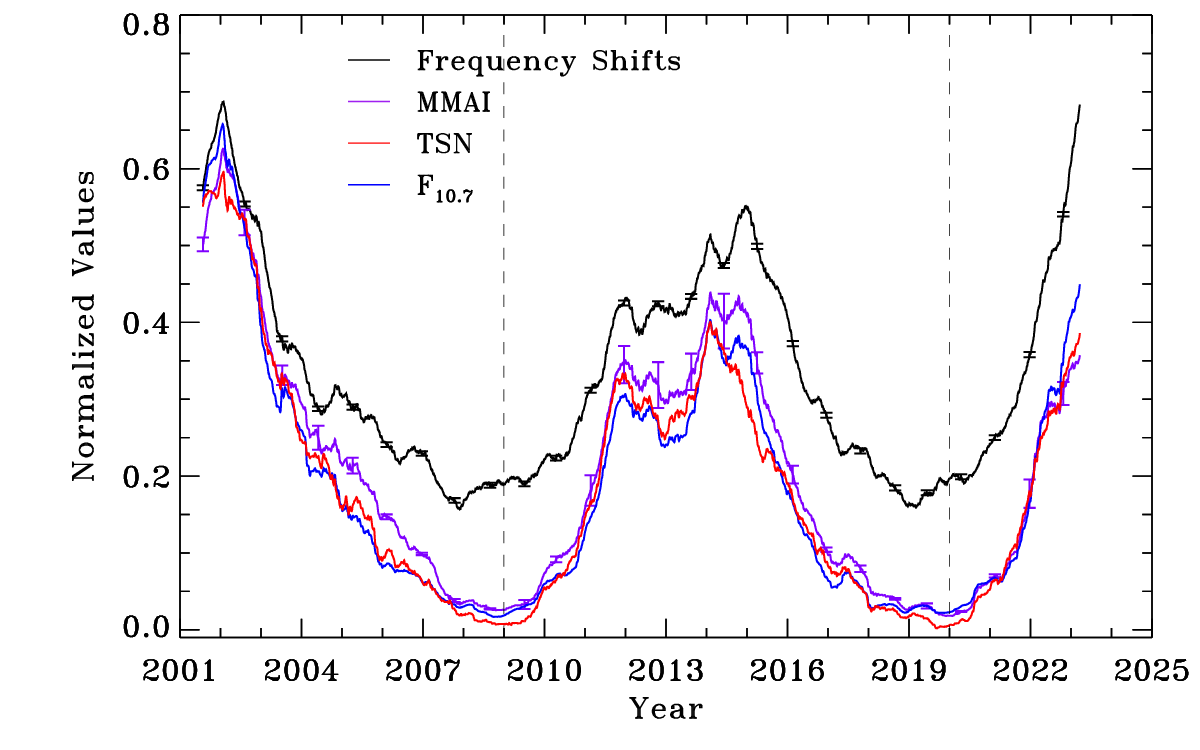}    
\caption{The temporal evolution of frequency shifts, $\delta\nu$, mean magnetic activity index, MMAI, total sunspot number, TSN,  and the 10.7 cm radio flux, $F_{10.7}$. All quantities are scaled to have values between 0 and 1.  We only show the smoothed values which are obtained  by applying a boxcar average over 181 data points.   The errors for $\delta\nu$ and MMAI are shown for every 250th point and are  multiplied by a factor of 5 to be visible on this plot. The dashed vertical lines have the same meaning as in Figure~\ref{dutytime}.    \label{fig:anuact}}
\end{figure}

\begin{figure}[ht!]
\centering
 \includegraphics[width=0.8\textwidth]{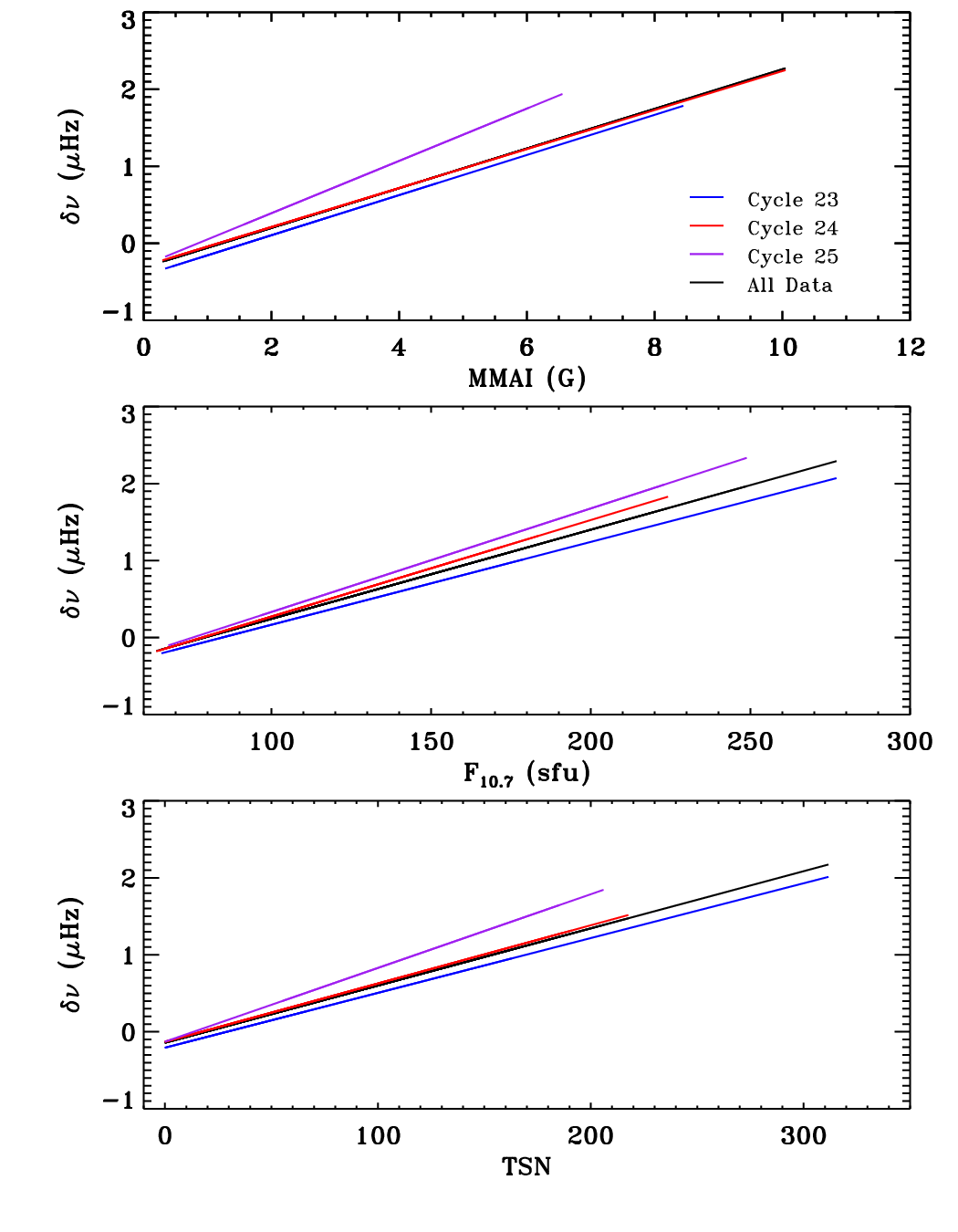}    
\caption{The lines show the result of linear regression fits obtained between $\delta\nu$ and different activity proxies;  bottom panel: total sunspot number, TSN,  middle panel: 10.7 cm radio flux, $F_{10.7},$ and upper panel: mean magnetic activity  index, MMAI.  The different colors represent different solar cycles and are indicated in the top panel. We do not plot the error bars since these are small but their values are tabulated in Table~\ref{T-tab1}. Note that the red and black lines in the upper panel are mostly overlapping. 
    \label{fig:aslopes}}
\end{figure}
\clearpage
\pagebreak
\newpage
\bibliography{Baird}    
\end{document}